\documentclass[lettersize,journal]{IEEEtran}
    \usepackage{makecell}
    \usepackage{array}
    \usepackage{graphicx,amssymb,amsmath}
    \usepackage{multicol}
    \usepackage[noadjust]{cite}
    \usepackage{setspace}
    \usepackage{subfigure}
    \usepackage{graphicx}
    \usepackage{float}
    \usepackage {url}
    \usepackage{stfloats}
    \usepackage{amsthm,pifont}
    \usepackage{flushend}
    \usepackage{cases,subeqnarray}
    \usepackage{bm,multirow,bigstrut}
    \usepackage{amsmath, amsthm, amssymb}
    \usepackage{textcomp}
    \usepackage{latexsym,bm}
    \usepackage{booktabs}
    \usepackage{xcolor}
    \usepackage{mathtools}
    \usepackage{dsfont}
    \usepackage{extarrows}
    \usepackage{epsfig}
    \usepackage{epsfig}
    \usepackage{epstopdf}
    \usepackage[noend]{algpseudocode}
    \usepackage{algorithmicx,algorithm}

    \theoremstyle{plain}

    \theoremstyle{plain}

    \newtheorem{them}{Theorem}

    \usepackage{amsmath}
    
    \newtheorem{lemma}{Lemma}
    
    \newcommand{\NAME}{\textsc{PerSF-SemCom}}
    \newcommand{\lizh}{\textcolor{black}}
    
    \IEEEoverridecommandlockouts
    
\begin{document}
    \title{{\color{black}Personalized Saliency in Task-Oriented Semantic Communications: Image Transmission and Performance Analysis}}
    \author{Jiawen Kang, Hongyang Du, Zonghang Li, Zehui Xiong, Shiyao Ma,   Dusit Niyato, \emph{Fellow, IEEE}, Yuan Li

\IEEEcompsocitemizethanks{
	Jiawen Kang is with Automation of School, Guangdong University of Technology, China.  (Email: kavinkang@gdut.edu.cn).
	
	Hongyang Du and Dusit Niyato are with School of Computer Science and Engineering, Nanyang Technological University, Singapore.
	
	Zonghang Li is with School of Information and Communication Engineering, University of Electronic Science and Technology of China, Chengdu, China. (Email: zhli@std.uestc.edu.cn).
	
	Zehui Xiong is with Pillar of Information Systems Technology and Design, Singapore University of Technology and Design, Singapore.
	
	Shiyao Ma is with College of Information and Communication Engineering, Dalian Minzu University.
	
	Yuan Li is with Shangdong University, China.
	} 
    }
    \maketitle
    \vspace{-1cm}
    \begin{abstract}
    Semantic communication, as a promising technology, has emerged to break through the Shannon limit, which is envisioned as the key enabler and  fundamental paradigm for future 6G networks and applications, e.g., smart healthcare. In this paper, we focus on UAV image-sensing-driven task-oriented semantic communications scenarios.  The majority of existing work has focused on designing advanced algorithms for high-performance semantic communication. However, the challenges, such as energy-hungry and efficiency-limited image retrieval manner, and semantic encoding without considering user personality, have not been explored yet. These challenges have hindered the widespread adoption of semantic communication. To address the above challenges, at the semantic level, we first design an energy-efficient task-oriented semantic communication framework with a triple-based {\color{black}scene graph} for image information. We then design a new personalized semantic encoder based on user interests to meet the requirements of personalized saliency. Moreover, at the communication level,  we study the effects of dynamic wireless fading channel on semantic transmission mathematically and thus design an optimal multi-user resource allocation scheme by using game theory.  Numerical results based on real-world datasets  clearly indicate that the proposed framework and schemes significantly enhance the personalization and anti-interference performance  of semantic communication,  and are also efficient to improve the communication quality of semantic communication services.
    
    \end{abstract}
    \begin{IEEEkeywords}
    Semantic communication, personalized saliency, resource allocation, unmanned aerial vehicle.
    \end{IEEEkeywords}
    \IEEEpeerreviewmaketitle
    \section{Introduction}
    \subsection{Background and Motivations}
    The fast-growing 6G communication technology  is enabling  the transition from serving people and things to supporting the ``Internet of Everything''~\cite{dang2020should,liu2020federated,giordani2020toward}. Specifically, 6G communication technology serves intelligent production and life through intelligent interconnection and collaborative symbiosis of human-machine-object, and actively promotes the construction of an inclusive and intelligent human society~\cite{you2021towards,zhang20196g,yang2022semantic}. In the 6G era, however, the traditional point-to-point information transmission communication system, that relies on the resource optimization of physical-layer dimension and the stable transmission protocol at the network layer, cannot meet the increasing requirements of  complex, diverse, and intelligent information transmission needs, e.g., supporting virtual reality, holographic projection and Metaverse applications~\cite{saad2019vision,xu2022full}. Therefore, it is essential to design  a new communication paradigm for efficient information transmission thus meeting the demands of  future communication.
    
    Fortunately, semantic communication~\cite{zhang2022toward}, as a new architecture that can integrate user needs and information semantic features into the communication process, is expected to become a new communication paradigm for the Internet of Everything in the future~\cite{strinati20216g,lan2021semantic,shi2021semantic}. Different from the traditional communication architectures, the semantic communication system aims to fundamentally solve the problems of cross-system, cross-protocol, cross-network, and cross-human-machine information transmission redundancy in traditional information-transmission based communication protocols. The ultimate goal of the semantic communication system is to efficiently transmit content-aware and semantic-related information in a task-oriented manner, and make the grand vision of ``Internet of Everything'' come true. In this context, to better serve the two core technical standards of semantics and validity in 6G networks, the Task-Oriented Semantic Communication Systems (TOSC) was designed with two parts: semantic reconstruction and goal execution~\cite{xie2021task}
    
    For semantic reconstruction, a semantic feature extractor is applied to extract the semantic features behind the data to be transmitted and reconstruct the semantic information at the receiver. For example, Zhu \textit{et al.} in \cite{zhou2021semantic} proposed an adaptive transformer to encode the semantic information and decode it at the receiver. For task-oriented applications, semantic communication aims to extract semantic information related to the decision goal of the receiver. Prior work generally focuses on image recognition scenarios and develops image classification-oriented semantic communication for improving recognition accuracy rather than image semantic reconstruction~\cite{kang2021task,ayan2022task}. 
    Since Unmanned Aerial Vehicle (UAV) sensing has been widely applied in various industries with 6G, in this paper, we focus on investigating UAV sensing-driven task-oriented semantic communication systems.
    
    Particularly, UAV-sensing-driven task-oriented semantic communication aims to provide users with cross-regional intelligent services with the help of UAV's image retrieval, image recognition, image transmission, and image coding features. UAV-sensing-driven Task-oriented Semantic Communication (UTSC) generally  serves a multi-demand, complex cross-modal intelligent task with multiple users. The UTSC is particularly suitable for emerging intelligent scenarios in daily life~\cite{zhang2022toward,yang2022semantic} and industry~\cite{xie2020lite} (e.g., smart agriculture~\cite{friha2021internet}). 
    {\color{black}The service mode of UTSC is realized by collecting task demands from users and using UAV sensing equipment to collect image data, and returning demand feedback to users through cloud servers in the form of semantic information. For example, in smart agriculture, different farmers (i.e., users) require UAVs to accomplish different tasks (e.g., monitoring and shooting). In this case, it is challenging for UAVs to accomplish the task goals of all users simultaneously. Furthermore, due to resource limitations and power supply problems, UAVs cannot hover in the air for a long time to complete the personalized user demands in turn.} Therefore, there exists following unique challenges when designing USTC systems:
      \begin{itemize}     
      \item \textit{\textbf{Energy-hungry and efficiency-limited image retrieval:}} Traditional communication manners generally send all the captured images to the users, while many of the images are not interesting/needed by the users. Therefore, such inefficient communication manners may cause large waste of communication resources of UAVs and consume  unnecessary UAV energy. A straightforward solution is to use keyword subscriptions, namely, UAVs push images of interest to users by extracting {\color{black}scene graph}s of sensing images that match the subscription words provided by users. However, this solution cannot execute well in the traditional communication manners under the scenarios of limited UAV energy or poor wireless channel. It is because that image packet drop and re-transmission bring large delay. Thus the inefficiency of wireless transmission makes it difficult to realize real-time push according to keyword subscriptions. To this end, it is urgently necessary to design an efficient and energy-efficient semantic-based real-time subscription method to improve subscription accuracy and wireless resource utilization.
    
      \item \textit{\textbf{Semantic encoding without personality:}} Existing work ignores the personalized needs of semantic communication for encoding retrieval results and does not set user preferences for the importance of semantically encoded values. In the face of large-scale semantic communication, due to the limited bandwidth resources between UAVs and users, semantic coding is prone to signal fading during transmission, resulting in the drop of important coding values (i.e., important information for users). Therefore, we need to design a personalized semantic coding value weight setting scheme, and design different resource allocation schemes for users with different interests to ensure that important information about user preferences can be efficiently conveyed.
    
      \item \textit{\textbf{Without insights between wireless fading channel and semantic communication:}} {\color{black}The semantic triplet drop probability is defined as the probability that the number of error bits in the triplet exceeds the error correction capability. However, existing works rarely consider the drop of encoded information caused by the physical environment during information transmission. A common approach is to use neural networks to model the wireless channel, which cannot help the system design with mathematical analysis~\cite{xie2021deep}. The problems about the multi-path effect, shadow effect, and co-channel interference for wireless transmission environment affect the communication quality and increase the semantic {\color{black}triplets drop} problem during transmission. Thereby, it is essential to study the joint optimization problem of the wireless fading channel and semantic {\color{black}triplet drop probability}, and also design a resource optimization scheme.}
    \end{itemize} 
    
    \subsection{Solutions and Contributions}
    To address the above challenges, in this paper, we propose an energy-efficient task-oriented semantic communication framework for 6G-enabled UAV image sensing scenarios. More specifically, in this framework, image information is modeled as a triple-based {\color{black}scene graph} to provide users with images that meet their preference requirements in an efficient retrieval manner. On this basis, we further execute the triplets by weight-encoding and use a personalized attention-based mechanism to implement differential weight encoding of triplets for important information according to user preferences. Moreover, for the UAV power allocation issues, we further consider the dynamic wireless fading effects on the semantic information transmission, and thus mathematically analyze the triplet drop probability. Based on the theoretical analysis results, we formulate a game theory model and design a multi-user resource allocation scheme to achieve efficient resource utilization and maximize the resource utility of UAVs.

    The key contributions are summarized as follows:
    \begin{itemize}
    \item Unlike traditional UAV-sensing communications that require all images to be transmitted, we propose an energy-efficient semantic communication-based framework that enables UAV only to transfer the selected interested images of the users, which is achieved by matching the user's query text with the semantic information of all images. 
    \item We design a novel personalized semantic encoder with the help of the user's subjective interest. After obtaining the semantic information, i.e., triples, from the image, the triplets of more interest to the user are given higher weights, thus ensuring the correct reception.
    \item We analyze the performance of semantic triplets transmission mathematically. We derive the exact expression for the semantic {\color{black}triplet drop probability} by considering the generalized fading channel model. From the derived expressions, we obtain insights into the wireless channel environment on the impact of semantic communication.
    \item Considering the resource limitation of UAVs and the requirements to send semantic information to multiple users, we propose a multi-user resource allocation scheme based on game theory, whose utility function of the retrieval task is used as the optimization objective for better resource utilization.
    \end{itemize}

The remainder of this paper is organized as follows. We first summarize the related work about Semantic Communication in Sec. \ref{sec-2}. We then introduce our system model in Sec. \ref{sec-3}. Then, we present the proposed semantic triplets transmission method from the communication level in Sec. \ref{sec-4}. Next, we elaborate on the proposed semantic communication framework in Sec. \ref{SemanticAnalysis}. Subsequently, we conduct a series of case studies and analyze the simulation results in Sec. \ref{sec-6}. Finally, we conclude this paper in Sec. \ref{sec-7}. We summarize the mathematical symbols and explanations in Table \ref{table:symbols}.

  
    \begin{table}
    \small
  \caption{Summary of Main Symbols.}
  \label{table:symbols}
  \centering
  \renewcommand{\arraystretch}{1.2}
  \begin{tabular}{p{40pt}<{\centering}|p{185pt}<{}}
    \hline
    \textbf{Symbol} & \textbf{Explanation} \\
    \hline\hline
     $m_{f}$ & The fading parameter for the $k_{\rm th}$ user in Fisher-Snedecor $\mathcal{F}$ fading model.\\
     \hline
    $m_{s}$ & The shadowing parameter for the $k_{\rm th}$ user in Fisher-Snedecor $\mathcal{F}$ fading model.\\
    \hline
    ${\bar z}$ & The average value of $\mathcal{F}$ random variables (RVs), i.e., $z$, for the $k_{\rm th}$ user in Fisher-Snedecor $\mathcal{F}$ fading model.\\
    \hline
        $\bm{a}^T$ &   Transpose of vector $\bm{a}$. \\
        \hline
    $\Gamma\!\left( \cdot \right) $ & Gamma function \cite[eq. (8.310.1)]{gradshteyn2007}. \\
    \hline
          $\Gamma\!\left( { \cdot , \cdot } \right)$ & Upper incomplete Gamma function \cite[eq. (8.350.2)]{gradshteyn2007}.\\
          \hline
    ${B\!\left( { \cdot , \cdot } \right)}$ & Beta function \cite[eq. (8.384.1)]{gradshteyn2007}.  \\
  \hline
    $F\left( { \cdot , \cdot ; \cdot ; \cdot } \right)$ & Gauss hypergeometric function \cite[eq. (9.111)]{gradshteyn2007}, which is also known as ${}_2{F_1}\left( { \cdot , \cdot ; \cdot ; \cdot } \right)$.  \\
    \hline
    $G \, \substack{ m, n \\ p, q}(\cdot)$ & Meijer's $G$-function \cite[eq. (9.301)]{gradshteyn2007}. \\
      \hline
    $H_{ \cdot  \cdot }^{ \cdot  \cdot }\left( { \cdot \left|  \cdot  \right.} \right)$ & Multivariate Fox's $H$-function \cite[eq. (A-1)]{mathai2009h}.\\
      \hline
    $G \, \substack{ \cdot, \cdot: \cdot, \cdot: \cdot, \cdot\\ \cdot, \cdot: \cdot, \cdot: \cdot, \cdot}(\cdot) $ & Bivariate Meijer's $G$-function \cite[eq. (1)]{sharma1974generating}.\\
    \hline
    $N_T$ & The number of antennas in the UAV \\
    \hline
    $x$ & The input image. \\
    \hline
    $K$ & The number of users. \\
    \hline
    $k$ & The identity of user. \\
    \hline
    $q^k$ & The personal query text of user $k$. \\
    \hline
    $\mathcal{T}$ & The Triplet Detection (TD) function. \\
    \hline
    $\mathcal{P}$ & The Personalized Saliency Prediction (PSP) function. \\
    \hline
    $\mathcal{C}$ & The crop function, used to crop a sub-image from $x$ according to the box coordinates. \\
    \hline
    $\tau$ & The set of (subject-relation-object) triplets. \\
    \hline
    $\tau_\mathrm{recv}^k$ & The triplets received by user $k$. \\
    \hline
    $H_{\mathrm{sub}}, H_{\mathrm{obj}}$ & The attention heatmaps of subject and object entities. \\
    \hline
    $B_{\mathrm{sub}}, B_{\mathrm{obj}}$ & The box coordinates of subject and object entities. \\
    \hline
    $p^k$ & The personalized triplet priority of user $k$. \\
    \hline
    $S^k$ & The personalized saliency heatmap of user $k$. \\
    \hline
    $\alpha$ & The coefficient for fusing attention heatmap and saliency heatmap. \\
    \hline
    $F^k_{\mathrm{sub}},F^k_{\mathrm{obj}}$ & The fused attention heatmaps of subject and object entities of user $k$. \\
    \hline
    $\tilde{F}^k_{\mathrm{sub}},\tilde{F}^k_{\mathrm{obj}}$ & The sub-heatmaps of subject and object entities of user $k$, cropped from $F^k_{\mathrm{sub}},F^k_{\mathrm{obj}}$. \\
    \hline
    $s$ & The match score between the triplets received and the personal query. \\
    \hline
  \end{tabular}
\end{table}

    \section{Related Work}\label{sec-2}
  
\subsection{Semantic Communication}
Semantic communication is a new communication paradigm, and it can transmit more information when the external environment is the same \cite{qin2021semantic}. The sender sends the semantic information extracted from the original information (such as images, text, and video) to the receiver in a semantic communication system and the receiver recovers the original information from the semantic information. By transmitting the semantic information, the semantic communication system can eliminate the unnecessary information from the original information to reduce communication overhead. The research on semantic communication is roughly divided into four parts: how to optimize semantic encoding, how to complete the goal-orient communications, how to protect semantic information privacy, and how to analyze the systems' performance.

\textbf{Semantic Encoding Optimization:} Semantic encoding is crucial for the semantic communication system because it determines the efficiency and effectiveness of information transmission. Xie et al. \cite{xie2021deep} proposed a DeepSC framework that is based on a Transformer for text transmission tasks, DeepSC can extract the semantic information from the original text under the interference of noise. Xie et al. \cite{xie2020lite} further considered the more realistic situation that designing a lightweight deep learning semantic communication system makes the model easier to deploy on IoT devices. The above work only focused on text-based semantic communication. Besides, images-based semantic communication has been proposed. Bourtsoulatze \cite{bourtsoulatze2019deep} proposed a semantic communication system that utilizes joint source and channel coding techniques to reduce communication overhead for image transmission tasks. Although the above work \cite{xie2021deep, xie2020lite, bourtsoulatze2019deep} proposed the semantic communication system to reduce the communication overhead, they only considered the transmission on single-domain, such as text-to-text or image-to-image. It is unrealistic in the real world. Therefore, we consider the transmission of multi-domain and multimodal in this paper. We design the framework in which the sender extracts the triplets with semantic information from the original images, and then the sender sends the triplets to the receiver. It can further reduce the communication overhead and be more practical.

\textbf{Task-oriented Communications:} The task-oriented communication is also important for the semantic communication system because it transmits different semantic information according to different needs. That is to say, task-oriented communication systems do not need to transmit data directly under corresponding circumstances. To this end, Farshbafan et al. \cite{farshbafan2021common} proposed a task-oriented semantic communication framework, which can transmit different semantic information in the different goals of the system. However, reference \cite{farshbafan2021common} only considered a single sender and receiver at the same time. Xie at el. \cite{xie2021task} proposed a task-oriented multi-user semantic communication system, the framework utilized different encoders and decoders to solve the multi-task and multi-user problems. Although the above work \cite{farshbafan2021common, xie2021task} proposed the task-oriented communication system to complete the corresponding goal, they ignored the individual differences in different users, i.e., other receivers will have different needs in the real world. Therefore, we consider the personalized saliency in our framework to adapt to different users' goals.

\textbf{Secure Semantic Encoding:} Since semantic information can reflect the real data distribution of users to a certain extent and is also vulnerable to privacy leakage in communication, we need to protect the semantic information transmitted by users. For example, Chen et al. \cite{chen2020joint} proposed a federated learning framework in semantic communication systems, which can protect the privacy of the system. Yang et al. \cite{yang2020energy} proposed a federated learning-based semantic communication framework to cope with the computing and communication overhead under protecting the privacy of the system. The above work illustrates that federated learning can be applied to wireless communication systems to protect the devices' data privacy. Therefore, Tong et al. \cite{tong2021federated} designed a wav2vec-based autoencoder federated semantic communication framework, and it can significantly reduce transmission error and heavy communication overhead. Due to the low efficiency of the current semantic communication system, we focus on how to extract semantic information effectively and effectively adapt to different goals. 

\textbf{Semantic Communications Performance Analysis: }In a wireless semantic communication system, the transmission performance of the system is negatively affected by multipath fading, that is, interference between different signals. Therefore, it is crucial to accurately model fading channels to better understand the performance impact of fading channels on semantic communication systems. Previous work focused on the performance analysis of wireless communication systems. For example, Yoo et al. \cite{yoo2017fisher} proposed the $\mathcal{F}$ distribution as fading model to analyze the performance of the semantic communication systems. Based on this, reference \cite{8638956} further explored a comprehensive performance analysis of the $\mathcal{F}$ composite fading channels in conventional wireless communication systems. Due to the rise of semantic communications, some researchers currently turned attention to studying the impact of fading channels on the performance of wireless semantic communication systems. Xie et al. \cite{9252948}  and Weng et al. \cite{9450827} explored the performance of their framework over Rayleigh and Rician channels in semantic communication systems. However, reference \cite{9252948, 9450827} only explored the communication performance in Rayleigh and Rician fading channels that are the traditional multipath fading models. {\color{black}Due to the Fisher-Snedecor $\mathcal{F}$ fading channel is a commonly used fading channel model in wireless communication systems, and it can cover the situation of classical fading channel analysis through changing parameters. Therefore, in this paper, we analyze the performance of our model by considering Fisher-Snedecor $\mathcal{F}$ channel model in wireless semantic communication and obtain insights into the wireless channel environment on the impact of semantic communication.}

\subsection{Personalized Saliency}
{\color{black}Personalized saliency means that different observers have different regions of interest in the same image. In recent years, personalized saliency has received a lot of attention in the computer vision community. Xu et al. \cite{xu2018personalized} proposed a CNN-based and personal information framework to predict a personalized saliency map. However, personalized information and images always change in the real world. Dodge et al. \cite{dodge2018visual} proposed a framework that combines the global scene information from all categories and the extracted local information to predict the saliency map. However, many categories are unknowable. Mahdi et al. \cite{mahdi2019deepfeat} utilized three CNN-based models to obtain the bottom-up and top-down deep features to predict a personalized saliency map. {\color{black} However, due to the storage capacity and the computing resources, it may be difficult to utilize this framework for extracting complex features when the number of users increases. Berkovsky et al. \cite{berkovsky2019detecting} proposed a framework for predicting personalized saliency based on eye tracking data and the framework needs to capture physiological responses, such as brain signals. However, there exist massive users in semantic communication systems and it is difficult to capture signals between multiple users, so the framework is difficult to adapt to multi-user situations. Moroto et al. \cite{moroto2021few} proposed the framework to extract personalized saliency maps (PSMs) through Gaussian process regression. However, considering the limitation of storage capacity, the receivers, i.e., UAVs, are difficult to store the PSMs for all users. Therefore, we convert PSMs into triples to further reduce memory overhead in this paper.} The above work may be unsuitable for semantic communication systems, especially for UAV scenarios. Therefore, we consider the situation of a multi-user semantic communication system and propose personalized saliency-based semantic communication.} 

In summary, we propose a personalized saliency-based task-oriented semantic communication system to cope with the above problems in this paper. Firstly, we predict the saliency heatmap of the user through the customized information and the image captured by the UAV. Meanwhile, UAV executes triplet detection \cite{cong2022reltr} to generate an attention heatmap from the image captured by the UAV. Secondly, UAV executes the attention fusion step to obtain the fused attention for each user and obtains the personalized triplet from the fused attention. The above steps complete the purpose of personalized saliency. Thirdly, UAV allocates multi-user power through triplet priority estimation and transmits the triplets by power allocated. Finally, the user obtains the match score between the received triplets and personal query and decides whether to download the image based on the matching score. The last steps complete the purpose of goal-orient semantic communication.

    \section{System Model}\label{sec-3}
    \begin{figure}[t]
    \centering
    \includegraphics[scale=0.55]{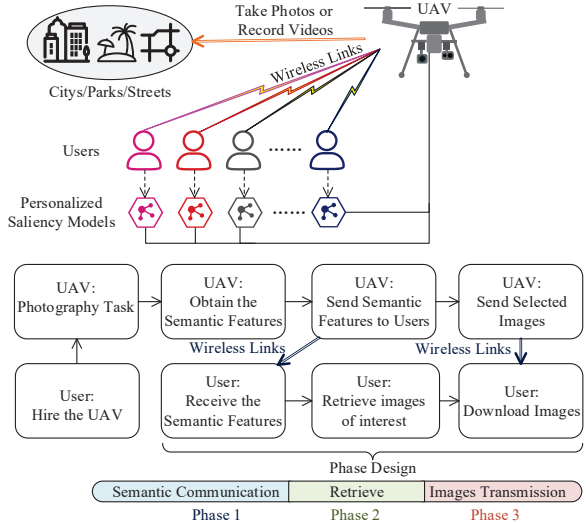}
    \caption{An illustration of the proposed task-oriented SemCom system model.}
    \label{fig:model}
    \end{figure}
    
    In this section, we first describe our proposed task-oriented SemCom system and then describe the metrics used to evaluate the effectiveness of the proposed design.
    
    \subsection{Task-oriented SemCom Design}
    When users hire a UAV for image acquisition, in many cases, not all the images taken by the UAV are what the user needs. Therefore, a retrieval task needs to be performed on all the images, i.e., the users input the text of the scene they want and get the corresponding images. However, considering that the energy of the UAV is limited and that one UAV may need to serve multiple users, performing all the users' retrieval tasks in the UAV will affect the quality of images and the UAV's endurance. Moreover, sending all the images to users and retrieving images on the user side also involve unnecessary energy consumption, i.e., users need to download some images that are not of interest.
    
   The development of semantic communication has given us a solution to solve the above problems. As shown in Fig. \ref{fig:model}, instead of transmitting all the images back to users after the photography task, the UAV transmits the semantic features (text format) of the images to users. The transmission of semantic features of images in text format requires few channel resources and is convenient for users to store. The users determine a query text according to their interests, match the images in the received semantic features, and then download the original image from the UAV. In the rest of this paper, we focus on the extraction and transmission of semantic information.
    
    \subsection{Optimization Problems Analysis}
    The UAV extracts the triplets that represent the corresponding image information from the captured images. If $ {{n_i}} $ triplets can be extracted from the $i_{\rm th}$ image, we have ${{\bm \tau} _i} = \left\{ {{\tau _{i,1}}, \ldots ,{\tau _{i,{n_i}}}} \right\} $, where ${{\bm \tau} _i}$ denotes the set of triplets from the $i_{\rm th}$ image. In Phase 1, the UAV transmits the triplets to the users. Note that users are not necessarily located close to each other, and the different interests of users will make them have different query texts. Therefore, we consider the TDMA scheme in Phase 1. In each time slot, i.e., $T_1$, the UAV uses all the antennas to serve one user and designs the beamforming vector accordingly. Let us consider an energy-constrained scenario, which means that
    \begin{equation}
    \sum\limits_{j \in {\cal K}} {{P_j}T_1}  < {W_A},
    \end{equation}
    where {\color{black}$\cal K$ is the users selected in this round}, ${P_j}$ is the transmit power for the $j_{\rm th}$ user and $W_A$ is the total energy.
    Thus, we need to solve two optimization problems:
    \begin{itemize}
    \item {\textbf{P1-power allocation among users:}} One user per time slot is served. Because the total energy of the UAV is limited, we need to determine how many resources each user can occupy, in the form of transmit power.
    \item {\textbf{P2-power allocation among one user's triplets:}} After determining the available power for each user, we also need to determine the power that should be allocated to each triplet.
    \end{itemize}
    
    Let $s_k$ denote the match score between the triplets received and the personal query of the $k_{\rm th}$ user, 
    \begin{equation}
{s_k} = \frac{{{N_{{\mathop{\rm in}\nolimits} ,k}}}}{{{N_{\rm rec}}}},
\end{equation}
which represents the proportion of the number of interested images of the $k_{\rm th}$ user obtained by matching, i.e., $ {{N_{{\mathop{\rm in}\nolimits} ,k}}} $, to the total number of images captured by UAV, i.e., $ {{N_{\rm rec}}} $.

Let $\tilde s_k$ denote the optimal match score, 
    \begin{equation}
{\tilde s_k} = \frac{{{N_{{\mathop{\rm truth}\nolimits} ,k}}}}{{{N_{\rm rec}}}},
\end{equation}
which represents the proportion of the number of interested images of the $k_{\rm th}$ user in truth (without considering the impact of {\color{black}triplets drop} that may be caused by wireless transmission), i.e., $ {{N_{\rm truth}}} $, to  $ {{N_{\rm rec}}} $. Then, we can use $\frac{s_k}{\tilde s_k}$ to represent the effectiveness of semantic communication.

    If P1 can be solved, we derive the power resources available to each user. Then, P2 for each user can be solved with the help of the personalized weights for the triplets. The solution for P2 is discussed in Section \ref{SemanticAnalysis}. Now we focus on P1. We consider the power allocation problem among users as cooperative bargaining. Thus, with the help of NBS, we need to maximize $ \prod\limits_{k = 1}^K {\frac{{{s_k}}}{{{{\tilde s}_k}}}}  $. Because ${\tilde s}_k$ represents the ground-of-truth which is only decided by the query text, the optimization problem can be transformed into
    \begin{equation}
    \mathop {\max }\limits_{{P_k}\left( {k = 1, \ldots ,K} \right)} \prod\limits_{k = 1}^K {{s_k}}.
    \end{equation}
    
    For a given query text of the $k_{\rm th}$ user, the $s_k$ is mainly affected by the number of received triplets. Due to the random nature of the wireless communication environment, the transmitted triplets may not be decoded correctly because of excessive error codes. Therefore, the higher transmit power should be allocated to the more important triplets, i.e., the triplets that are of more interest to the user, to guarantee error-free transmission.
    
    \subsection{SINR Analysis}\label{section:network-analysis}
    We consider a set of ${\cal K}=1,\ldots,K$ user, where each user has their own preference. One UAV performs the photography task and needs to transmit the semantic features of the image to $K$ users. To obtain considerable array gains and improve the channel quality, we consider the UAV is equipped with $N_T$ antennas. The UAV is hovering above the ground $K$ users. The horizontal coordinate of the $k_{\rm th}$ ground device is assumed to be $ {u_{{k}}} = \left( {{x_{{k}}},{y_{{k}}},0} \right) $, $ {k} = \left( {1, \ldots ,K} \right) $, while the UAV is hovering at a fixed altitude $z_u$ with the coordinate $ {u_{\rm u}} = \left( {{x_{\rm u}},{y_{\rm u}},{z_{\rm u}}} \right) $. Thus, the distance between the $k_{\rm th}$ user and the UAV can be expressed as $ {D_{{\rm u}{k}}} = \sqrt {{{\left\| {{u_{{k}}} - {u_{\rm{u}}}} \right\|}^2}} $. Let $\alpha_k$ denote the path loss exponents of the UAV-$k_{\rm th}$ user link.
    
    We denote the channel vector from the UAV to the $k_{\rm th}$ user as ${\bm h}_k \in {\mathbb C}^{1\times K}$. By adopting the linear beamforming, the data symbol {\color{black}{$t_k$}} intended for user $k$ is multiplied with the beamformer ${\bm w}_k \in {\mathbb C}^{K\times 1}$. {\color{black}$N_{Ik}$ paths of interferes are assumed to be present at the $k_{\rm th}$ user, where $I$ means the abbreviation of interference, which is used to distinguish symbols. Each of the $j_{\rm th}$ user interfering signals has an average transmit power $P_{Ik}$. {\color{black}{${t_{I,k,j}}$}} is the $j_{\rm th}$ interfering symbol.}
    Accordingly, because TDMA is used, the received signal at the $k_{\rm th}$ user is given by\footnote{Note that the large scale fading of the interference signal is considered in the mean value of $h_{I,k,j}$.}
    \begin{equation}
    {\color{black}{r_k}} = \sqrt{{P_k}{D_{{\rm u}{k}}^{-\alpha_k}}}{{\bf{h}}_k}{{\bf{w}}_k}{x_k} + {{P_{Ik}}} \sum\limits_{j = 1}^{{N_{Ik}}} {\sqrt {{h}_{I,k,j}} {\color{black}{{t_{I,k,j}}}} } + {n_k},
    \end{equation}
    where $n_k \in {\cal{CN}}\left( {0,{\sigma ^2}} \right)$ is the noise, $P_k$ is the transmit power for the $k_{\rm th}$ user. The SINR at the $k_{\rm th}$ user can be expressed as
    \begin{equation}
    {\gamma _k} = \frac{{{P_k}D_{{\rm{u}}k}^{ - {\alpha _k}}{{\left\| {{{\bf{h}}_k}{{\bf{w}}_k}} \right\|}^2}}}{{{\sigma ^2} + {P_{Ik}} \sum\limits_{j = 1}^{{N_{Ik}}} {h_{I,k,j}^2} }}.
    \end{equation}
    With the help of maximum ratio transmission~\cite{lo1999maximum}, the optimal beamforming vector can be expressed as $ {{\bf{w}}_k} = \frac{{{{\bf{h}}_k}^T_1}}{{\left\| {{{\bf{h}}_k}} \right\|}} $. Thus, we have ${\left\| {{{\bf{h}}_k}{{\bf{w}}_k}} \right\|^2} = \sum\limits_{j = 1}^{{N_T}} {h_{k,j}^2}$.
    
    \subsection{Channel Model}
    {\color{black}The small scale fading of UAV-$k_{\rm th}$ user link is modeled as the Fisher-Snedecor $\mathcal{F}$ fading distribution. The Fisher-Snedecor $\mathcal{F}$ composite fading model assumes that small-scale variations follow the Nakagami-$m$ distribution and shadowing follows the inverse Nakagami-$m$ distribution. Channel measurements at 5.8 GHz have demonstrated that the Fisher-Snedecor $\mathcal{F}$ fading model fits experimental results better than the KG fading model both in line-of-sight and non-LOS scenarios~\cite{yoo2017fisher}.}
    
 Thus, ${\left\| {{{\bf{h}}_k}{{\bf{w}}_k}} \right\|^2}$ follows the distribution of sum of $N_T$ Fisher-Snedecor $\mathcal{F}$ RVs \cite{du2020sum}. However, the PDF and CDF of ${\left\| {{{\bf{h}}_k}{{\bf{w}}_k}} \right\|^2}$ is in terms of Multivariate Fox's $H$-function \cite[eq. (A-1)]{mathai2009h}, which is hard to provide insights. Considering that the Fisher-Snedecor $\mathcal{F}$ RV is defined as the ratio of two Gamma RVs and the sum of Gamma RVs still follows the Gamma distribution, we can use the single Fisher-Snedecor $\mathcal{F}$ distribution to approximate the distribution of the sum of Fisher-Snedecor $\mathcal{F}$ RVs \cite{du2020sum}.
    
    Let $Z \triangleq {\left\| {{{\bf{h}}_k}{{\bf{w}}_k}} \right\|^2} \sim\mathcal{F}\left( {{m_{fk}},{m_{sk}},\bar z_{k}} \right) $, the PDF and CDF of $Z$ are given as \cite[eq. (6)]{yoo2019comprehensive} and \cite[eq. (12)]{yoo2019comprehensive}, respectively. 
    
    We consider the interference signals follow the Rayleigh distribution, i.e., $ {h_{I,k,j}} \sim {\rm Rayleigh} \left( {\eta _{k}}\right)$. Let $Y \triangleq \sum\limits_{j = 1}^{{N_{Ik}}} {h_{I,k,j}^2}$. Because the sum of $N_{Ik}$ i.i.d. Rayleigh-fading signals have a Nakagami-$m$ distributed signal amplitude with $m=N_{Ik}$, the PDF and CDF expressions of $\lVert\mathbf{h}_{k}\rVert^2$ can be written as \cite{nakagami1960m}.
    \begin{equation}\label{PDFnakagemi}
    f_{Y}\left( y \right) = \frac{{{y^{N_{Ik} - 1}}}}{{\eta _{k}^{N_{Ik}}{\Gamma\!\left( {{N_{Ik}}} \right)}}}\exp\!\left( { - \frac{y}{{{\eta _{k}}}}} \right),
    \end{equation}
    and
    \begin{equation}\label{CDFnakagemi}
    F_{Y}\left( y \right) = \frac{{\Gamma\!\left( {{N_{Ik}},\frac{y}{{{\eta _{k}}}}} \right)}}{{\Gamma\!\left( {N_{Ik}} \right)}},
    \end{equation}
    where $ {{\eta _{k}}}={{{\mathbb E}\left[ h_{k}^2 \right]}}$, and ${\mathbb E}\left[ \cdot \right]$ denotes expectation.
    We then derive the PDF and CDF of the SINR, $ {\gamma _k} = \frac{{{P_k}D_{{\rm{u}}k}^{ {\alpha _k}}Z}}{{{\sigma ^2} + {P_{Ik}}Y}} $.
    
    \begin{them}\label{LemmaPDFF}
    The PDF and CDF of the SINR, $ {\gamma _k} = \frac{{{P_k}D_{{\rm{u}}k}^{ {\alpha _k}}Z}}{{{\sigma ^2} + {P_{Ik}}Y}} $, can be derived as \eqref{PDFFINAL} and \eqref{CDFFINAL}, respectively, shown at the bottom of the next page, where ${\Lambda _k} \triangleq \frac{{{P_k}D_{{\rm{u}}k}^{ - {\alpha _k}}\left( {{m_{sk}} - 1} \right){{\bar z}_k}}}{{{m_{fk}}}}$, $ {\Theta _1} = \left( {1 - {N_{Ik}}:\left\{ { - 1, - 1,0} \right\}} \right) $, and $ {\Theta _2} = \left( {1 - {m_{fk}} - {m_{sk}}:\left\{ {0, - 1, - 1} \right\}} \right) $.
    \newcounter{mycount}
    \begin{figure*}[b]
    \normalsize
    \setcounter{mycount}{\value{equation}}
    \hrulefill
    \vspace*{4pt}
    \begin{align}\label{PDFFINAL}
    {f_{{\gamma _k}}}\left( \gamma  \right) &= \frac{{{\Lambda _k}^{{m_{sk}}}{{\left( {{\sigma ^2}} \right)}^{{N_{Ik}} + {m_{fk}}}}{\gamma ^{{m_{fk}} - 1}}}}{{{{\left( {{P_{Ik}}{\eta _k}} \right)}^{{N_{Ik}}}}{{\left( {{\Lambda _k} + \gamma {\sigma ^2}} \right)}^{{m_{fk}} + {m_{sk}}}}{\Gamma\!\left( {{N_{Ik}}} \right)}\Gamma\!\left( {{m_{sk}}} \right)\Gamma\!\left( {{m_{fk}}} \right)}}
    \notag\\ &\times
    G_{0,0:0,2:1,2}^{1,0:2,0:1,1}\left( {\left. {\begin{array}{*{20}{c}}
    {{N_{Ik}}}\\
    - 
    \end{array}} \right|\left. {\begin{array}{*{20}{c}}
    - \\
    {{m_{sk}} - {N_{Ik}},0}
    \end{array}} \right|\begin{array}{*{20}{c}}
    {1 - {m_{fk}} - {m_{sk}}}\\
    {0,1 - {m_{sk}}}
    \end{array}\left| {\frac{{{\sigma ^2}}}{{{P_{Ik}}{\eta _k}}},\frac{{ - {\Lambda _k}}}{{{\Lambda _k} + \gamma {\sigma ^2}}}} \right.} \right)
    \end{align}
    \begin{align}\label{CDFFINAL}
    &{F_{{\gamma _k}}}\left( \gamma  \right){\rm{ = }}\frac{{{\Lambda _k}^{ - {m_{fk}}}{{\left( {{\sigma ^2}} \right)}^{{N_{Ik}} + {m_{fk}}}}{\gamma ^{{m_{fk}}}}}}{{{{\left( {{P_{Ik}}{\eta _k}} \right)}^{{N_{Ik}}}}{\Gamma\!\left( {{N_{Ik}}} \right)}\Gamma\!\left( {{m_{sk}}} \right)\Gamma\!\left( {{m_{fk}}} \right)}}
    \notag\\ &\times\!\!
    H_{2,0:2.0;2,0;2,1}^{0,2:0,2;0,1;1,1}\!\!\!\left(\!\!\!\! {\left. {\begin{array}{*{20}{c}}
    {{{ {{P_{Ik}}{\eta _k}} }}{\sigma ^{-2}}}\\
    { - 1}\\
    {{\sigma ^{-2}}{\Lambda _k}\gamma^{ - 1} }
    \end{array}} \!\!\right|\!\!\!\begin{array}{*{20}{c}}
    {{{\Theta _1},{\Theta _2}}}\\
    - 
    \end{array}\left[ {\begin{array}{*{20}{c}}
    {\left( {1 - {m_{sk}} + {N_{Ik}},1} \right)\left( {1,1} \right)}\\
    - 
    \end{array}} \right]\left[ {\begin{array}{*{20}{c}}
    {\left( {1,1} \right)\left( {{m_{sk}},1} \right)}\\
    - 
    \end{array}} \right]\left[ {\begin{array}{*{20}{c}}
    {\left( {1,1} \right)\left( {1 + {m_{fk}},1} \right)}\\
    {\left( {{m_{sk}},1} \right)}
    \end{array}} \right]} \right)
    \end{align}
    \setcounter{equation}{\value{mycount}}
    \end{figure*}
    \addtocounter{equation}{2}
    \begin{IEEEproof}
    Please refer to Appendix \ref{LemmaPDFFA}.
    \end{IEEEproof}
    \end{them}
    
    \subsection{Approximation Analysis}    
    Although the previously derived PDF and CDF are obtained in closed-form, it is hard to bring valuable insights if we derive performance expressions with the help of \eqref{PDFFINAL} and \eqref{CDFFINAL}. Therefore, in the following, we derive the accurate approximate CDF by presenting an approximate solution to a complex mathematical integral equation. Moreover, we analyze the approximate CDF in the high-SNDR regime and verify our derived results by numerical analysis.
    
    \subsubsection{Accurate Approximation}
    Let us consider the integration as follows:
    \begin{equation}
    I_A = \int_a^\infty  {{x^b}{{\left( {x - a} \right)}^c}\exp\!\left( { \frac{{x - a}}{d}} \right){F}\left( {\alpha ,\beta ,\varepsilon ; \rho x} \right){\rm{d}}x}.
    \end{equation}
    Note that $I_A$ has not been studied in any mathematical integral theory book or website such as \cite{gradshteyn2007,web}. It is difficult, if not impossible, to obtain the closed solution of $I_A$. Here we derive the accurate approximate solution of $I_A$ in the Lemma \ref{2lemma}.
    \begin{lemma}\label{2lemma}
    An accurate approximation of $I_A$ when $\rho$ is small can be expressed as
    \begin{equation}\label{APPFINAL}
    {I_A} = \exp\!\left( {\frac{a}{d}} \right)\!\!\frac{{\Gamma\!\left( \varepsilon  \right){d^{b + c + 1}}}}{{\Gamma\!\left( \alpha  \right)\Gamma\!\left( \beta  \right)}}G_{3,2}^{1,3}\left(\!\! {d\rho \left|\!\! {\begin{array}{*{20}{c}}
    {1 \!-\! \beta , - b - c,1\! -\! \alpha }\\
    {1 \!-\! \varepsilon }
    \end{array}} \right.} \!\!\!\right).
    \end{equation}
    \begin{IEEEproof}
    Please refer to Appendix \ref{2lemmaA}.
    \end{IEEEproof}
    \end{lemma}
    With the help of Lemma \ref{2lemma}, we can re-derive the CDF of $\gamma_k$ as
    \begin{them}\label{2them}
    An accurate CDF of $\gamma_k$ can be obtained as
    \begin{align}
    {F_{{\gamma _k}}}\left( \gamma  \right) &= \frac{1}{{\Gamma\!\left( {{N_{Ik}}} \right)\Gamma\!\left( {{m_{sk}}} \right)\Gamma\!\left( {{m_{fk}}} \right)}}\exp\!\left( {\frac{{{\sigma ^2}}}{{{P_{Ik}}{\eta _k}}}} \right)
    \notag\\&\times
    G_{3,2}^{1,3}\left( {\frac{{\gamma {P_{Ik}}{\eta _k}}}{{{\Lambda _k}}}\left| {\begin{array}{*{20}{c}}
    {1 - {m_{sk}},1 - {N_{Ik}},1}\\
    {{m_{fk}},0}
    \end{array}} \right.} \right).
    \end{align}
    \end{them}
    \subsubsection{High-SNDR approximation}
    In the following, we analyze asymptotic CDF in the high-SNDR regime.
    \begin{them}
    The CDF of $\gamma_k$ can be approximated in the high transmit power regime as
    \begin{align}\label{CDFHIGH}
{F_{{\gamma _k}}}\left( \gamma  \right) =& \frac{{\Gamma\!\left( {{m_{fk}} + {m_{sk}}} \right)\Gamma\!\left( {{N_{Ik}} + {m_{fk}}} \right)}}{{\Gamma\!\left( {{N_{Ik}}} \right)\Gamma\!\left( {{m_{sk}}} \right)\Gamma\!\left( {{m_{fk}} + 1} \right)}}
\notag\\&\times
\exp\!\left( {\frac{{{\sigma ^2}}}{{{P_{Ik}}{\eta _k}}}} \right){\left( {\frac{{\gamma {P_{Ik}}{\eta _k}}}{{{\Lambda _k}}}} \right)^{{m_{fk}}}}.
\end{align}
    \end{them}
    \subsubsection{Verification and Insights}
\begin{figure}[t]
    \centering
    \includegraphics[width=.5\textwidth]{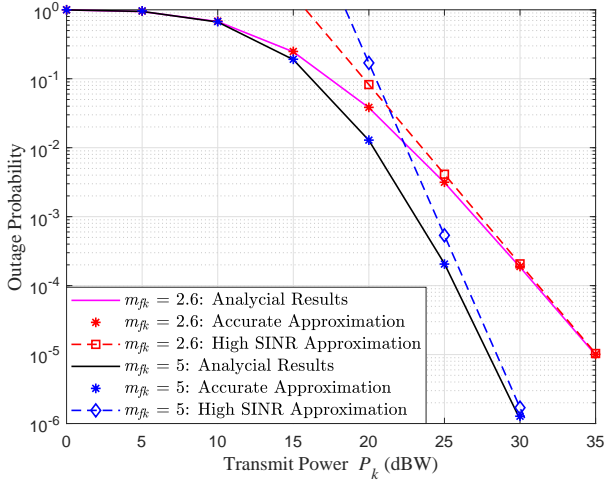}
    \caption{The outage probability versus the transmit power, with different multi-path fading parameter $m_{fk}$.}
    \label{fig:ApproOP}
    \end{figure}

   We use the OP to verify the derived CDF expression, \eqref{CDFFINAL}, and two approximate expressions, \eqref{APPFINAL} and \eqref{CDFHIGH}. The outage probability (OP) is defined as the probability that the SINR falls below a given outage threshold, i.e., $OP_{\rm k} = {\mathbb P} ( \gamma_k < \gamma_{\rm th}) = F_{\gamma_k}(\gamma_{\rm th})$. As shown in Fig. \ref{fig:ApproOP}, we study the OP versus the transmit power, with $D_{\rm uk}=1.5$ ${\rm m}$, $\alpha_k=2$, $m_{sk}=5$, $m_{fk}=2.6$, $\bar z_k=-1$ ${\rm dB}$, $\sigma^2 = 1$ ${\rm W}$, $\eta_k=0.4$, $P_{Ik}=5$ ${\rm W}$, $N_{Ik}=3$, and different values of $m_{fk}$. We can observe that the accurate approximate expression \eqref{APPFINAL} matches almost exactly with the closed-form analytic expression \eqref{CDFFINAL}. In the high-SINR regime, e.g., when $P_k$ is larger than $25$ ${\rm dBW}$, the values obtained from the high-SINR approximate expression \eqref{CDFHIGH} are close to that obtained from analytic expression. Furthermore, from \eqref{CDFHIGH}, we can observe that the multi-path fading parameter $m_{fk}$, instead of the shadowing parameter $m_{sk}$, determines the slope of the OP with decreasing transmit power. In Fig. \ref{fig:ApproOP}, it is shown that the larger the $m_{fk}$ is, the more rapidly the OP decreases with the increase of transmit power. 
    
    \section{Communication Level: Triplet Drop Probability}\label{sec-4}
    {\color{black}We encode the semantic information of the image into triplets. Due to the instability of wireless transmission, the receiver cannot guarantee the perfect receiving of the semantic triplets transmitted by the UAV. Therefore, we analyze the impact of the wireless environment on the semantic triplets transmission.} Suppose a triplet is encoded with bit length $D_T$, and that the use of bit error correction codes allows for at most $D_E$ error bits.
    \subsection{Bit Error Probability}
    {\color{black}{Under a variety of modulation formats, the BEP, $E_k$, can be expressed as \cite[eq. (13)]{zhang2017new}
    \begin{equation}\label{eqber}
    E_k = \int_0^\infty  \frac{{\Gamma\!\left( {{ \color{black}{\lambda} _2},{\color{black}{\lambda} _1}\gamma } \right)}}{{2\Gamma\!\left( {{\color{black}{\lambda} _2}} \right)}}{f_{\gamma_k}\left( \gamma  \right){\rm{d}}\gamma },
    \end{equation}
    where $ {{{\Gamma\!\left( {{\color{black}{\lambda} _2},{\color{black}{\lambda} _1}\gamma } \right)}}/{{2\Gamma\!\left( {{\color{black}{\lambda} _2}} \right)}}} $is the conditional bit error probability, ${\color{black}{\lambda} _1} $ and ${\color{black}{\lambda} _2}$ are modulation-specific parameters which have different values under different modulation and detection schemes \cite{zhang2017new}.}}

    \begin{them}\label{berthm}
    The BEP of the $k_{\rm th}$ user can be derived as
    \begin{align}\label{BERFINAL}
&{E_k} = \frac{{{\Gamma ^{ - 1}}\left( {{m_{sk}}} \right){\Gamma ^{ - 1}}\left( {{m_{fk}}} \right)}}{{2\Gamma\!\left( {{\color{black}{\lambda} _2}} \right)\Gamma\!\left( {{N_{Ik}}} \right)}}\exp\!\left( {\frac{{{\sigma ^2}}}{{{P_{I,k}}{\eta _k}}}} \right)
\notag\\&\times
G_{4,2}^{1,4}\left( {\frac{{{P_{I,k}}{\eta _k}}}{{{\color{black}{\lambda} _1}{\Lambda _k}}}\left| {\begin{array}{*{20}{c}}
{1 - {m_{sk}},1 - {N_{Ik}},1,1 - {\color{black}{\lambda} _2}}\\
{{m_{fk}},0}
\end{array}} \right.} \right).
\end{align}

    \begin{IEEEproof}
    Please refer to appendix \ref{berthmA}.
    \end{IEEEproof}
    \end{them}
    
    \subsection{Triplet Drop Probability}
    We consider that the BEP of $k_{\rm th}$ user is $E_k$. The TDP $P_k$ can be expressed as
    \begin{equation}
    P_k = \sum\limits_{j = D_E + 1}^{D_T} {{E_k}^j{{\left( {1-{E_k}} \right)}^{{D_T}-j}}},
    \end{equation}
    which can be calculated with the help of \eqref{eqber}.
    
 \section{Semantic Level: Personalized Saliency Fused Semantic Communication Framework}\label{SemanticAnalysis}
 
 \begin{figure*}[t]
 	\centering
 	\includegraphics[width=\textwidth]{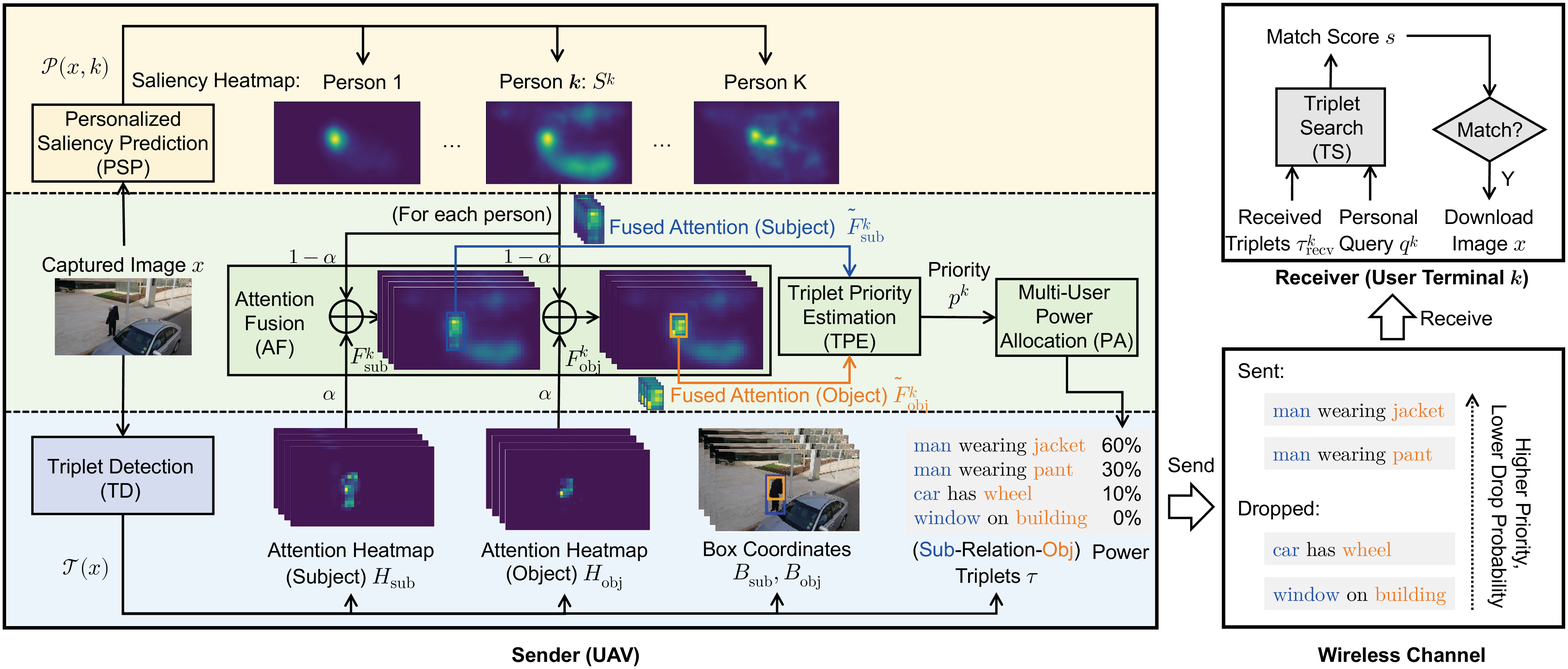}
 	\caption{An overview of the Personalized Saliency Fused Semantic Communication Framework (\NAME).}
 	\label{fig:sender_receiver_architecture}
 \end{figure*}
 
 In applications such as real-time sensing of UAV aerial photography, the autonomous patrol UAV cruises along the specified trajectory and transmits aerial pictures back to users (e.g., monitoring centers and photographers). In such a wireless communication environment with fading channels, transmitting all aerial images to all users is expensive, which is impractical in reality and limits the application of UAV aerial photography subscription. In this work, we propose that users have the right to specify which type of pictures they prefer, then the UAV only needs to transmit a small number of specific images to each user. For example, user A only needs to download pictures of ``man wearing a jacket", and user B prefers pictures of ``bus on the street." These texts are called personal queries, based on which the user decides whether to download an image from the UAV. Does the problem come from \textit{how do personal users know they need a given picture without downloading the original picture?} This paper addresses this problem by proposing a Personalized Saliency Fused Semantic Communication framework (\NAME), which leverages a Triplet Detector to infer a fixed-size set of (subject-relation-object) triplets for each picture, and transmits these tiny-size triplets to users for triplet matching, in which only the matching images need to be downloaded. Besides, to prevent the triplets concerned by personalized users from being lost in the harsh wireless competitive environment, we quantify the priority of triplets and propose a transmission power allocation strategy, under which triplets with higher priority have more transmission power and therefore have a lower probability of being dropped.
 
 \subsection{OA-SemCom: A Fully Objective Approach}
 {\color{black}As shown in Figure \ref{fig:sender_receiver_architecture}, the architecture mainly contains Triplet Detection (TD), Personalized Saliency Prediction (PSP), Attention Fusion (AF), Triplet Priority Estimation (TPE), and Power Allocation (PA) on the UAV, and a Triplet Search (TS) module on the user terminal. Whenever the UAV captures a picture $x$, TD $\mathcal{T}(x)$ infers a fixed-size set of (subject-relation-object) triplets, i.e., the semantic information, $\tau$, from $x$, and respectively the attention heatmaps $H_{\mathrm{sub}}, H_{\mathrm{obj}}$ and box coordinates $B_{\mathrm{sub}}, B_{\mathrm{obj}}$ of their subject and object entities. 
 	\lizh{Specifically, the pretrained RelTR~\cite{cong2022reltr} model is used as the kernel of the TD module, which builds an encoder-decoder architecture like Transformer~\cite{vaswani2017attention}, where the encoder infers the visual feature context that is then used by the decoder for triplet inference. However, the TD module can also be implemented using any other pretrained scene graph generation technique. For example, two-stage approaches employ Fast/Faster R-CNN~\cite{girshick2015fast}\cite{ren2015faster} to extract object features, and then apply scene graph generation~\cite{xu2017scene}\cite{zellers2018neural} for graph inference. The alternative is one-stage approaches such as FCSGG~\cite{liu2021fully} and the RelTR we use, which predict objects and their relations concurrently, in an end-to-end fashion, and are thus more lightweight and faster. Regardless of which technique is used to implement TD, the general formula $\mathcal{T}(x)$ for TD obeys
 		\begin{equation}
 			H_{\mathrm{sub}},H_{\mathrm{obj}}, B_{\mathrm{sub}},B_{\mathrm{obj}},\tau\gets \mathcal{T}(x),
 		\end{equation}
 		where $H_{\mathrm{sub}},H_{\mathrm{obj}}$ are objective attention heatmaps of subjects and objects, $B_{\mathrm{sub}},B_{\mathrm{obj}}$ are bounding boxes of subjects and objects, respectively, and $\tau$ is the prediction set of (subject-relation-object) triplets.}}
 
 After this step, a naive idea would be to transmit the triplet set $\tau$ to all users to match their personal queries $q^k (\forall k\in[1,K])$. This naive approach (named Naive-SemCom) faces the problem of key triplets being dropped due to intense competition among multiple users for scarce wireless channel resources. For example, the user $k$ has the personal query $q^k=\it{man\ wearing\ jacket}$, but unfortunately, the packet of the key triplet $\it{man\ wearing\ jacket}$ is dropped in the wireless channel, then the current image will fail to match the personal query. The reason is that the wireless channel treats packets of all triplets as equally important, making the key triplets drop with equal probability as other triplets. 
 
 To this end, we propose prioritizing triplets for each user and allocating more transmission power to triplets with higher priority to ensure that key triplets are successfully delivered to the user terminal. To achieve this, one challenge should be addressed, that is, \textit{how to prioritize triplets and identify the key triplets for personalized users?} As a benchmark, we can use the product of maximum values of the subject and object attention heatmaps $H_{\mathrm{sub}}, H_{\mathrm{obj}}$ to obtain the triplet priority $p$ in an objective view, namely, $p=\max(\mathcal{C}(H_{\mathrm{sub}}, B_{\mathrm{sub}}))\otimes\max(\mathcal{C}(H_{\mathrm{obj}}, B_{\mathrm{obj}}))$, where $\mathcal{C}(H_{\mathrm{sub}}, B_{\mathrm{sub}})$ is a cropping function that crops the attention sub-image from $H_{\mathrm{sub}}$ according to the box coordinates $B_{\mathrm{sub}}$. Note that the cropping function $\mathcal{C}$ and $\max$ in this section are channel-wise operations, and ``$\otimes$" is an element-wise multiplier. We refer to this benchmarking approach as Objective Attention-Based Semantic Communication (OA-SemCom), which, as the name suggests, only considers the objective global attention of the image itself, and its triplet priority is common to all users. However, users have personalized saliency, and their triplets should be prioritized differently than others, as are key triplets, but OA-SemCom fails to capture the personalization in subjective saliency among different users.

 \subsection{\NAME: Achieve Personalization Through Fused Saliency}
 To address the above issues, we develop a \NAME~framework that integrates objective visual attention from RelTR and subjective visual attention from a personalized saliency prediction module, namely PSP, to assist the UAV in personalized priority estimation for each user. \lizh{In our implementation, the PSP module $\mathcal{P}(x,k)$ leverages a pretrained fully convolutional encoder-decoder network structure\cite{kroner2020contextual} to predict personal saliency heatmap $S^k$ for user $k$,}
 \begin{equation}
 	S^k\gets\mathcal{P}(x,k).
 \end{equation}
 These heatmaps show the things that different users are most interested in when looking at the same image: the users' subjective saliency distribution. For example, some users may only follow persons, while others also follow cars. 
 
 \lizh{The encoder uses VGG16~\cite{simonyan2014very} without pooling layers as the backbone, followed by an ASPP module to capture multi-scale visual information, and then the decoder restores the original image resolution by stacking convolution and up-sampling layers. Since how to achieve personalization in PSP is not the focus of this study, for simplicity, we use a plain but effective method to achieve personalization, which is to train user saliency models separately on different user datasets to obtain personalized behaviors. Please note that similar to the TD module, the PSP module is also a replaceable plugin that can be replaced with other pretrained personalized saliency models, depending on the researcher's preference. For instance, treat each user's prediction task separately and use a multi-task model to train a personalized saliency model for each user~\cite{xu2018personalized}, or train a meta-learning model that can quickly adapt to new personalized tasks~\cite{luo2022few}.}
 
 
 Then in Figure \ref{fig:sender_receiver_architecture}, we take a person $k$ as an example to illustrate how to fuse objective attention and subjective saliency. In the AF module, the attention heatmaps $H_{\mathrm{sub}}, H_{\mathrm{obj}}$ and the saliency heatmap $S^k$ are first normalized, and then fused by weighted sum, where $\alpha\in[0,1]$ is the fusion coefficient, ``$\oplus$" is the broadcast add operator, and $\mathrm{norm}$ is a global normalizer,
 \begin{align}
 	F^k_{\mathrm{sub}} &= \alpha\cdot \mathrm{norm}(S^k) \oplus (1-\alpha)\cdot \mathrm{norm}(H_{\mathrm{sub}}), \label{eq:fuse_sub}\\
 	F^k_{\mathrm{obj}} &= \alpha\cdot \mathrm{norm}(S^k) \oplus (1-\alpha)\cdot \mathrm{norm}(H_{\mathrm{obj}}). \label{eq:fuse_obj}
 \end{align}
 In order to accurately locate the attention heatmaps of the subject and object entities, the fused attention heatmaps $F^k_{\mathrm{sub}},F^k_{\mathrm{obj}}$ should be cropped according to the box coordinates $B_{\mathrm{sub}},B_{\mathrm{obj}}$ to obtain the sub-heatmaps $\tilde{F}^k_{\mathrm{sub}},\tilde{F}^k_{\mathrm{obj}}$ of the subject and object entities. These sub-heatmaps are then fed into the TPE module for priority estimation, where their maxima are multiplied and used as the triplet priorities $p^k$,
 \begin{gather}
 	\tilde{F}^k_{\mathrm{sub}} = \mathcal{C}(F^k_{\mathrm{sub}},B_{\mathrm{sub}}), \label{eq:crop_sub}\\
 	\tilde{F}^k_{\mathrm{obj}} = \mathcal{C}(F^k_{\mathrm{obj}},B_{\mathrm{obj}}), \label{eq:crop_obj}\\
 	p^k = \max(\tilde{F}^k_{\mathrm{sub}})\otimes\max(\tilde{F}^k_{\mathrm{obj}}). \label{eq:priority}
 \end{gather}
 Finally, the PA module allocates transmission power for triplets according to personalized priorities $p^k$. The key triplets with higher priority will be allocated more transmission power, which makes them less likely to be dropped while traversing the wireless channel.
 
 On the receiver side (i.e., the user terminal $k$), the TS module uses the received triplets $\tau_\mathrm{recv}^k$ to match the personal query $q^k$ and calculate a match score $s$. Here, we recommend two types of TS modules: Accurate Mode (TS-AM) and Fuzzy Mode (TS-FM). TS-AM aims to find the same received triplet as the personal query, and it returns a match score of 1 if found and 0 otherwise. TS-AM is preferred if the user's personal query is forced to meet the (subject-relation-object) format. However, if the personal query is free text, TS-FM could be better because it can return a match score at the semantic level (e.g., HEM\cite{lu2020deep}). Once a matching score $s$ is obtained, the user can decide whether to download the current image $x$. In our implementation, we use TS-AM by default because semantic sentence matching is not the focus of this work, and only images with matching scores $s=1$ will be downloaded. We summarize the pseudo code of the proposed \NAME~in Algorithm \ref{alg:persf-semcom}.
 
 \begin{algorithm}[t]
 	\caption{\NAME~(Main)}\label{alg:persf-semcom} 
 	\begin{algorithmic}[1] 
 		\Require Captured image $x$ on UAV, user identity $k$.
 		\Ensure Match score $s$, downloaded image $x$ on user $k$.
 		\Procedure{UAV-Send}{$x,k$}
 		\State TD detects triplets $\tau$ and their attention heatmaps $H_{\mathrm{sub}},H_{\mathrm{obj}}$ and box coordinates $B_{\mathrm{sub}},B_{\mathrm{obj}}$: $$H_{\mathrm{sub}},H_{\mathrm{obj}}, B_{\mathrm{sub}},B_{\mathrm{obj}},\tau\gets \mathcal{T}(x);$$
 		\State PSP predicts the personalized saliency heatmap $S^k$ for user $k$: $S^k\gets\mathcal{P}(x,k)$;
 		\State AF fuses the objective attention $H_{\mathrm{sub}},H_{\mathrm{obj}}$ and the subjective saliency $S^k$ by Eqs. \eqref{eq:fuse_sub}-\eqref{eq:fuse_obj};
 		\State TPE calculates the priority $p^k$ of triplets $\tau$ using the fused attention heatmaps $F^k_{\mathrm{sub}},F^k_{\mathrm{obj}}$ by Eqs. \eqref{eq:crop_sub}-\eqref{eq:priority};
 		\State PA allocates transmission power to triplets $\tau$ according to their priority $p^k$ using RCGA\cite{herrera1998tackling};
 		\State UAV sends triplets $\tau$ to user $k$'s terminal;
 		\EndProcedure
 		\vspace{0.2cm}
 		\Procedure{User-Receive}{$\tau_{\mathrm{recv}}^k$}
 		\State TS calculates the match score $s$ between received triplets $\tau_{\mathrm{recv}}^k$ and personal query $q^k$ via TS-AM/FM;
 		\If{$s$ is greater than a user-specified threshold}
 		User $k$ downloads current image $x$;
 		\EndIf
 		\Return Match score $s$ and downloaded image $x$;
 		\EndProcedure
 	\end{algorithmic} 
 \end{algorithm}
 
 \section{Numerical Results}\label{sec-6}
 \subsection{Environment Setup}
\begin{figure}[t]
 	\centering
 	\includegraphics[width=.45\textwidth]{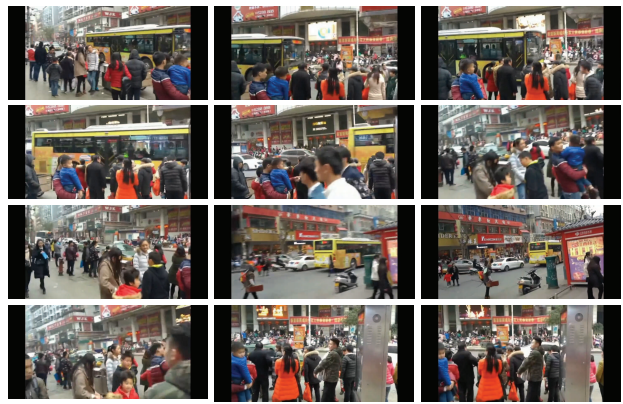}
 	\caption{A partial overview of our STREET dataset.}
 	\label{fig:street_dataset}
 \end{figure}

 The experimental platform is built on a generic Ubuntu 18.04 system with Intel(R) Xeon(R) E5-2678 CPU and 4 Geforce RTX 2080 TI GPUs. \lizh{The RelTR model is adopted as the core of the TD module, which has been pretrained on the Visual Genome (VG) dataset~\cite{krishna2017visual}, and has a top-50 recall rate of 25.2 on the scene graph detection metric, with the corresponding mean value being 8.5\footnote{The pretrained model is available at: \url{https://github.com/yrcong/RelTR}}. The VG dataset contains 108k images with 150 objects and 50 relationship categories. For the PSP module, we use the saliency prediction model in literature~\cite{kroner2020contextual} pretrained on 3 visual attention datasets, respectively, including SALICON~\cite{jiang2015salicon}, MIT1003~\cite{Judd_2009} and DUT-OMRON~\cite{yang2013saliency}, to simulate the visual saliency discrepancy of 3 users\footnote{The pretrained model is available at: \url{https://github.com/alexanderkroner/saliency}}.} The reason this works is that researchers collect these datasets from different perspectives, which can be regarded as real-world experiences and environments of different persons, thus presenting independent personalities. The validation dataset used in our experiments is real-world video frames downloaded from YouTube\footnote{The video is available at: https://www.youtube.com/watch?v=RPZ3xWy70IE}. We framed this video at an interval of 5 frames and obtained a dataset of 59 images, which we named ``STREET". In Figure \ref{fig:street_dataset}, we visualize a portion of the STREET dataset. Each user's personal queries and their experience datasets are summarized in Table \ref{table:user-info}.
 
 {\color{black}If not otherwise specified, the parameters of the small-scale fading model and the system parameters such as power and transmission distance are shown in Table \ref{table:channel-config} following common parameter settings in the literature \cite{zhang2020dual,du2019distribution}.}

 \begin{table}
 	\caption{Channel Configuration Parameters for 3 users.}
 	\small
 	\label{table:channel-config}
 	\centering
 	\renewcommand{\arraystretch}{1.2}
 	\begin{tabular}{c|c|c|c}
 		\hline
 		\textbf{User ID} & \textbf{1} & \textbf{2} & \textbf{3} \\
 		\hline\hline
 		Fading parameter $m_{f}$ & 2 & 2 & 5 \\
 		\hline
 		Shadowing parameter $m_{s}$ & 2 & 4 & 2 \\
 		\hline
 		Signal amplitude decrease $\bar{z}$ $({\rm dB})$ & \multicolumn{3}{c}{-3} \\
 		\hline
 		Distance $D_{\mathrm{u}k}$ $({\rm m})$ & \multicolumn{3}{c}{10} \\
 		\hline
 		Number of antennas $N_T$ & \multicolumn{3}{c}{3} \\
 		\hline
 		Paths of interferes $N_{I,k}$ & \multicolumn{3}{c}{2} \\
 		\hline
 		Interference power $P_{I,k,j}$ $({\rm W})$ & \multicolumn{3}{c}{2} \\
 		\hline
 		$\eta_k$ $({\rm dB})$ & \multicolumn{3}{c}{-3} \\
 		\hline
 		Noise $n_k$ $({\rm W})$ & \multicolumn{3}{c}{1} \\
 		\hline
 		$\tau_1$ & \multicolumn{3}{c}{1} \\
 		\hline
 		$\tau_2$ & \multicolumn{3}{c}{0.5} \\
 		\hline
 		Path loss exponent $\alpha_k$ & \multicolumn{3}{c}{2} \\
 		\hline
 	\end{tabular}
 \end{table}
 
 Two benchmark approaches are used for performance comparison, that is, Naive-SemCom and OA-SemCom. Naive-SemCom aims to allocate transmit power when transmitting to each user equally, and the triplets of each user are also allocated with equal power. In other words, Naive-SemCom has no awareness of the importance of semantic triples. Instead, OA-SemCom quantifies the priority of triplets and uses these priorities to schedule transmit power for each user and triplet. However, OA-SemCom only considers the objective attention obtained by TD but ignores the user's subjective attention and personality. To address this issue, our \NAME~is proposed, and PSP, AF modules are introduced to fuse the objective and subjective attention. We set the fusion coefficient $\alpha=0.2$ and the transmit power $P=3000$ by default and calculated the average of each user's match score on all images as the user's score. Each experiment was repeated 5 times, and the average results are shown.
 
 \begin{table}
 	\caption{users' identities, experience datasets and personal queries.}
 	\label{table:user-info}
 	\small
 	\centering
 	\renewcommand{\arraystretch}{1}
 	\begin{tabular}{c|c|c}
 		\hline
 		\textbf{Identity} & \textbf{Dataset} & \textbf{Personal Queries} \\
 		\hline\hline
 		User 1 & SALICON & woman has hair \\
 		\hline
 		User 2 & MIT1003 & sign on building \\
 		\hline
 		User 3 & DUT-OMRON & woman wearing shirt \\
 		\hline
 	\end{tabular}
 \end{table}
 
 \subsection{Results and Analysis}
 
\begin{figure}[t]
 	\centering
 	\includegraphics[width=.46\textwidth]{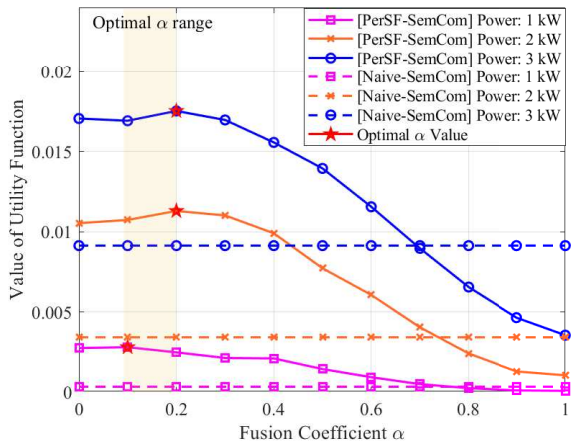}
 	\caption{The value curves of utility function over different fusion coefficient $\alpha$ and transmit power $P$.}
 	\label{fig:target_vs_alpha}
 \end{figure}

 \textbf{Effectiveness of \NAME~over $\alpha$ and $P$.}
 We first show the effectiveness of the proposed \NAME. To find the optimal hyperparameters, we increase the fusion coefficient $\alpha$ from 0 to 1 and the total transmits power $P$ from 1kW to 3kW. The utility value (i.e., the product of all user scores) curves are shown in Figure \ref{fig:target_vs_alpha}. Please note that OA-SemCom corresponds to our \NAME~with $\alpha=1.0$. Benefiting from the introduction of subjective attention, \NAME~outperforms Naive-SemCom in most cases, for example, when $\alpha\in[0,0.6]$ and $P\in[1\mathrm{kW}, \mathrm{3kW}]$. The curves of \NAME~first increase to the optimal and then rapidly degrade. The optimal utility value usually occurs when $\alpha$ is between 0.1 and 0.2; that is, the objective attention should contribute 10\%-20\% to the fused attention, while the subjective attention contributes 80\%-90\%. After this, the \NAME~performance gradually deteriorated with the withdrawal of subjective attention, especially when $\alpha=1.0$, \NAME~degenerates to OA-SemCom, which performs even worse than Naive-SemCom. The above results show that the introduction of appropriate subjective attention can significantly enhance semantic communication's personalization and anti-interference ability and demonstrates the effectiveness of the proposed \NAME. In subsequent experiments, we use the recommended $\alpha=0.2$ as the default setting.
 
\begin{figure}[t]
 	\centering
 	\includegraphics[width=0.46\textwidth]{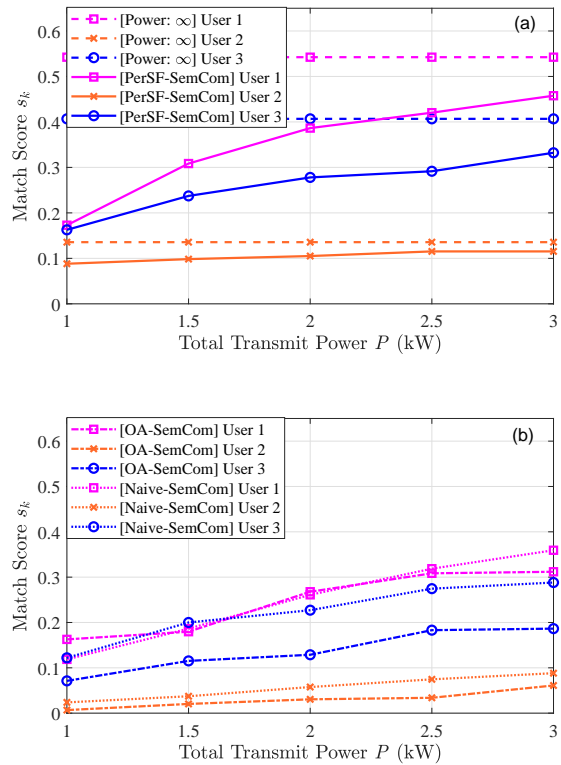}
 	\caption{The score curves for each user over different total transmit power $P$.}
 	\label{fig:score_vs_power}
 \end{figure}

 \textbf{The effects of transmit power $P$.} To explore the optimality gap over different transmit powers $P$, we increase $P$ from 1kW to 3kW and illustrate the gap between \NAME~and the theoretical optimal in Figure \ref{fig:score_vs_power}. The top dashed lines represent the theoretical upper bounds of each user's score. They use infinite transmit power, so no triplet packets get dropped. Users have different upper bounds because of their personality divergence. As expected, the optimality gap of \NAME~becomes smaller as the transmit power $P$ increases because higher transmit power effectively reduces the probability of packet loss. Compared with Naive-SemCom, when $P=3\mathrm{kW}$, \NAME~reduces the optimality gap by 54\%, 57\%, 37\% on the three users, respectively, which shows a significant improvement in the accuracy of the UAV aerial imagery subscription service. However, OA-SemCom, which only considers objective attention, widens the optimality gap by 26\%, 57\%, and 86\%. These results prove the necessity to incorporate personal attention.
 
\begin{figure}[t]
 	\centering
 	\includegraphics[width=.5\textwidth]{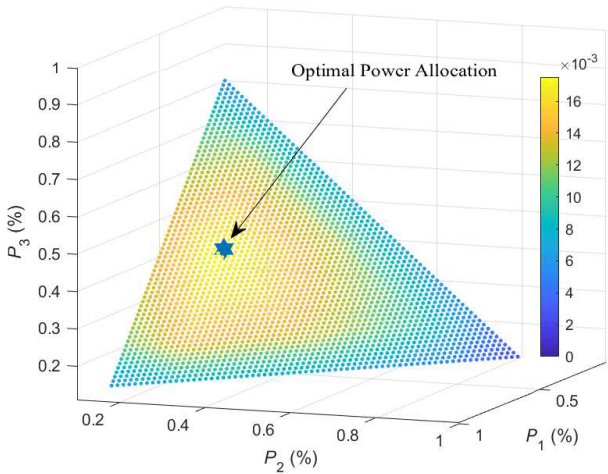}
 	\caption{The utility function surface under different power allocations.}
 	\label{fig:utility_value_vs_power}
 \end{figure}

 \textbf{The effects of power allocation.} Intuitively, users with poor channel conditions should be allocated more transmission power. In this experiment, the channel condition ranking of users 1-3 is user 1 $<$ user 3 $<$ user 2. We use the Real-Coded Genetic Algorithm (RCGA)\cite{herrera1998tackling} to solve the NBS problem and allocate power among users, and then proportionally distribute power among triplets according to priority. RCGA uses the population size of 50, the mutation probability of 0.001, and the maximum iteration of 20. Given the transmit power $P=3\mathrm{kW}$ and 3 users, we traverse all possible settings of full power allocation (i.e., the sum of the proportions of the power allocated to each user is 1) and visualize their utility function surface in Figure \ref{fig:utility_value_vs_power}. RCGA found the best power allocation per user to be (40.6\%, 18.4\%, 40.0\%), with the score (0.46, 0.12, 0.33) per user and the utility value 0.0175, which outperforms Naive-SemCom with the power allocation (33.3\%, 33.3\%, 33.3\%), the scores (0.36, 0.09, 0.29) and the utility value 0.0091. The resulting power allocation strategy also supports our intuitive idea that more power should be allocated to users with poor channel conditions rather than simply evenly allocated.
 
\begin{figure}[t]
 	\centering
 	\includegraphics[width=.46\textwidth]{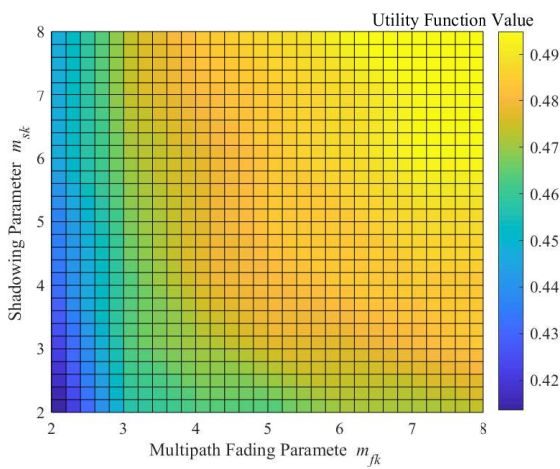}
 	\caption{Utility function values under different multi-path fading parameter $m_{fk}$ and shadow fading parameter $m_{sk}$.}
 	\label{fig:mms2}
 \end{figure}

 \textbf{The effects of small-scale channel conditions $m_{sk},m_{fk}$.} Given the transmission power of user $k$ is 1kW, we investigate the effect of different values of multi-path fading parameter $m_{fk}$ and shading parameter $m_{sk}$ on the utility function values. As shown in Fig.~\ref{fig:mms2}, when the values of both $m_{fk}$ and $m_{sk}$ are large, which means the multipath effect is weak and there is less shading, the value of the utility function is large. A more interesting insight is that an increase in $m_{fk}$ leads to a faster increase in the utility function value compared to the same increase in $m_{sk}$. This suggests that, at the wireless channel level, the goal-oriented semantic communication system that we studied in this paper is more affected by the BER increase due to the multi-path effect, instead of shadowing.
 
 \begin{figure}[t]
 	\centering
 	\includegraphics[width=.46\textwidth]{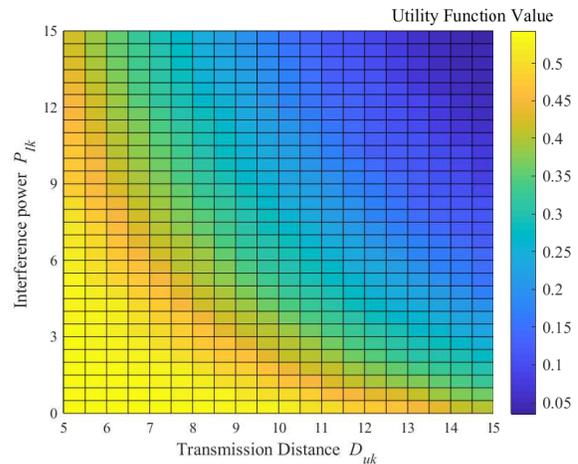}
 	\caption{Utility function values under different transmission distance $D_{\rm uk}$ and interference power $P_{Ik}$.}
 	\label{fig:pd2}
 \end{figure}
 \textbf{The effects of large-scale channel conditions $D_{\mathrm{u},k},P_{Ik}$.} Given the transmission power of user $k$ is 1kW, we investigate the effect of transmission distance and interference power on the value of the utility function. In Fig.~\ref{fig:pd2}, we can observe that, when the interference power is small, i.e., $P_{Ik}<1$ ${\rm W}$, the increase in the transmission distance does not result in a significant decrease in the value of the utility function. However, in the high interference regime, e.g., when $P_{Ik}>10$ ${\rm W}$, every 5 ${\rm m}$ increase in transmission distance results in an about $58\%$ decrease in the value of the utility function. Therefore, when there is substantial interference in the environment, we need to reduce the transmission distance by adjusting the trajectory of the UAV to ensure the quality of the semantic communication services.

 \begin{figure}[t]
 	\centering
 	\includegraphics[width=.36\textwidth]{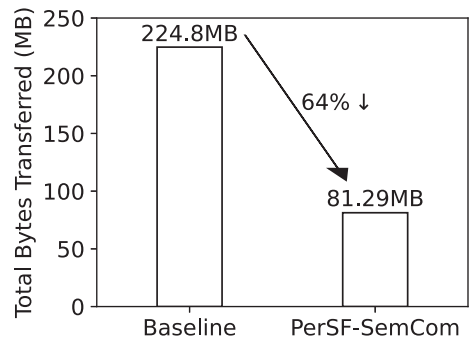}
 	\caption{Comparison of total transferred bytes of images.}
 	\label{fig:comm_cost}
 \end{figure}
 \lizh{\textbf{The effects of reducing communication overhead.} Considering a vanilla approach that the UAV sends a total of 59 images to 3 subscribers, each image is 1.27MB in size and the total bytes transferred is 224.8MB. As an improvement, \NAME~selectively sends fewer images (64 in total) to specified subscribers, before that 873 triplets were sent with a negligible additional communication cost of 10.24KB. Then, the UAV only needs to send 81.29MB in size, which reduces the communication cost by 64\% and thus saves the power consumption during transmission.}

    \section{Conclusions}\label{sec-7}
  In this paper, we have focused on the semantic communication personalization and resource allocation optimization issues for personalized saliency-based task-oriented semantic communication  in UAV image sensing scenarios. We have first presented an energy-efficient task-oriented semantic communication framework with an efficient image retrieval manner based on a triple-based {\color{black}scene graph}. To ensure personalized semantic communication, we have designed a personalized attention-based mechanism to realize differential weight encoding of triplets for important information according to user preferences. Furthermore, we have analyzed mathematically the effects of wireless fading channels on semantic communication and proposed a game-based model for a multi-user resource allocation scheme to achieve efficient utilization of UAV resources. We evaluate performance of the proposed framework and schemes on real-world datasets. The numerical results have confirmed that the proposed framework and schemes can realize personalized semantic communication and significantly enhance the UAV resource utilization.

    \begin{appendices}
    \section{Proof of Lemma \ref{LemmaPDFF}}\label{LemmaPDFFA}
    \renewcommand{\theequation}{A-\arabic{equation}}
    \setcounter{equation}{0}
    \subsection{Proof of PDF}
    Let $ X \triangleq {\sigma ^2} + {P_{Ik}}Y $ and $ U \triangleq {P_k}D_{{\rm{u}}k}^{ - {\alpha _k}}Z $. Thus, we have $ {\gamma _k} = \frac{U}{X} $. The PDF of ${\gamma _k}$ can be expressed as
    \begin{equation}\label{faeojo;}
    {f_{{\gamma _k}}}\left( \gamma  \right) = \int_0^\infty  {x{f_U}\left( {\gamma x} \right){f_X}\left( x \right){\rm{d}}x}.
    \end{equation}
    Substituting \eqref{PDFnakagemi} and \cite[eq. (6)]{yoo2019comprehensive} into \eqref{faeojo;}, we have 
    \begin{equation}\label{PDFDE}
    {f_{{\gamma _k}}}\!\left( \gamma  \right)\! = \!\frac{{{m_{fk}}^{{m_{fk}}}{{\left( {{m_{sk}} 1} \right)}^{{m_{sk}}}}{{\bar z}_k}^{{m_{sk}}}{\gamma ^{{m_{fk}} 1}}}}{{{{\left( {{P_{Ik}}{\eta _k}} \right)}^{{N_{Ik}}}}\Gamma\!\left( K \right){{\left( {{P_k}D_{{\rm{u}}k}^{ {\alpha _k}}} \right)}^{{m_{fk}}}}B\!\left( {{m_{fk}},{m_{sk}}} \right)}}I_1,
    \end{equation}
    where 
    \begin{equation}
    I_1 = \int_{{\sigma ^2}}^\infty  {\frac{{{x^{{m_{fk}}}}{{\left( {x {\sigma ^2}} \right)}^{{N_{Ik}} 1}}\exp\!\left( { \frac{{x {\sigma ^2}}}{{{P_{Ik}}{\eta _k}}}} \right)}}{{{{\left( {\frac{{{m_{fk}}\gamma x}}{{{P_k}D_{{\rm{u}}k}^{ {\alpha _k}}}} + \left( {{m_{sk}} 1} \right){{\bar z}_k}} \right)}^{{m_{fk}} + {m_{sk}}}}}}{\rm{d}}x}.
    \end{equation}
    With the help of \cite[eq. (9.301)]{gradshteyn2007} and \cite[eq. (01.03.26.0004.01)]{web}, we can express the $exp(\cdot)$ function in terms of the Mellin-Barnes integral form as
    \begin{align}\label{expextend}
    &\exp\!\left( { - \frac{{x - {\sigma ^2}}}{{{P_{Ik}}{\eta _k}}}} \right)  = G_{0,1}^{1,0}\left( {\left. {\frac{{x - {\sigma ^2}}}{{{P_{Ik}}{\eta _k}}}} \right|0} \right) 
    \notag\\
    & = \frac{1}{{2\pi i}}\int_{{{\cal L}_1}} {\Gamma\!\left( { - {s_1}} \right)} {\left( {\frac{{x - {\sigma ^2}}}{{{P_{Ik}}{\eta _k}}}} \right)^{{s_1}}}{\rm{d}}{s_1}.
    \end{align}
    Substituting \eqref{expextend} into $I_1$, we obtain that
    \begin{align}
    &{I_1} = \frac{1}{{2\pi i}}\int_{{{\cal L}_1}} {\Gamma\!\left( { - {s_1}} \right)} {\left( {\frac{1}{{{P_{Ik}}{\eta _k}}}} \right)^{{s_1}}}
    \notag\\&\times\!\!
    \int_0^\infty \!\!\! {\frac{{{x^{{s_1} + {N_{Ik}} - 1}}{{\left( {x\! +\! {\sigma ^2}} \right)}^{{m_{fk}}}}}{\rm{d}}x{\rm{d}}{s_1}}{{{{\left(\! {\frac{{{m_{fk}}\gamma x}}{{{P_k}D_{{\rm{u}}k}^{ - {\alpha _k}}}} \!+\! \frac{{{m_{fk}}\gamma {\sigma ^2}}}{{{P_k}D_{{\rm{u}}k}^{ - {\alpha _k}}}} \!+ \!\left(\! {{m_{sk}} \!-\! 1} \!\right){{\bar z}_k}} \!\right)}^{{m_{fk}} + {m_{sk}}}}}}} .
    \end{align}
    Let ${\Lambda _k} \triangleq \frac{{{P_k}D_{{\rm{u}}k}^{ - {\alpha _k}}\left( {{m_{sk}} - 1} \right){{\bar z}_k}}}{{{m_{fk}}}}$. With the help of \cite[eq. (3.197.1)]{gradshteyn2007}, the integration part in $I_1$ can be solved. Thus, we can re-write $I_1$ as
    \begin{align}
    &{I_1} = \frac{1}{{2\pi i}}{\left( {\frac{{{P_k}D_{{\rm{u}}k}^{ - {\alpha _k}}}}{{{m_{fk}}\gamma }}} \right)^{{m_{fk}} + {m_{sk}}}}\int_{\cal L} {\Gamma\!\left( { - {s_1}} \right)} 
    \notag\\&\times\!
    {\left( {\frac{1}{{{P_{Ik}}{\eta _k}}}} \right)^{{s_1}}}{\left( {\frac{{{P_k}D_{{\rm{u}}k}^{ - {\alpha _k}}\left( {{m_{sk}} - 1} \right){{\bar z}_k}}}{{{m_{fk}}\gamma }} + {\sigma ^2}} \right)^{ - {m_{fk}} - {m_{sk}}}}
    \notag\\&\times\!
    {\sigma ^{2{s_1} + 2{N_{Ik}} + 2{m_{fk}}}}B\!\left( {{s_1}\! + \!{N_{Ik}},{m_{sk}} - {s_1} - {N_{Ik}}} \right)
    \notag\\&\times\!
    F\!\left(\! {{m_{fk}}\! +\! {m_{sk}},{s_1} \!+\! {N_{Ik}};{m_{sk}};1 \!-\! \frac{{{\sigma ^2}}}{{\frac{{{\Lambda _k}}}{\gamma } + {\sigma ^2}}}} \!\right){\rm{d}}{s_1}.
    \end{align}
    Using \cite[eq. (9.113)]{gradshteyn2007} and \cite[eq. (8.384.1)]{gradshteyn2007}, we can further express $I_1$ as
    \begin{align}
    &{I_1} = {\left( {\frac{1}{{2\pi i}}} \right)^{\rm{2}}}{\left( {\frac{{{P_k}D_{{\rm{u}}k}^{ - {\alpha _k}}}}{{{m_{fk}}\left( {{\Lambda _k} + \gamma {\sigma ^2}} \right)}}} \right)^{{m_{fk}} + {m_{sk}}}}\frac{{{{\left( {{\sigma ^2}} \right)}^{{N_{Ik}} + {m_{fk}}}}}}{{\Gamma\!\left( {{m_{fk}} + {m_{sk}}} \right)}}
    \notag\\&\times\!
    \int_{{{\cal L}_1}} {\int_{{{\cal L}_2}} {\frac{{\Gamma\!\left( {{m_{sk}} \!-\! {s_1}\!- \!{N_{Ik}}} \right)\Gamma\!\left( { - {s_1}} \right)\Gamma\! \left( {{m_{fk}} \!+\! {m_{sk}} \!+ \!{s_2}} \right)}}{{\Gamma\!\left( {{m_{sk}} \!+\! {s_2}} \right){\Gamma ^{ - 1}}\left( { - {s_2}} \right){\Gamma ^{ - 1}}\left( {{s_1} \!+ \!{N_{Ik}} \!+ \!{s_2}} \right)}}} } 
    \notag\\&\times\!
    {\left( {\frac{{{\sigma ^2}}}{{{P_{Ik}}{\eta _k}}}} \right)^{{s_1}}}{\left( {\frac{{ - {\Lambda _k}}}{{{\Lambda _k} + \gamma {\sigma ^2}}}} \right)^{{s_2}}}{\rm{d}}{s_2}{\rm{d}}{s_1}.
    \end{align}
    
    Therefore, substituting $I_1$ into \eqref{PDFDE}, using the definition of Bivariate Meijer's $G$-function \cite[eq. (1)]{sharma1974generating}, we can derive \eqref{PDFFINAL} to complete the proof.
    
    \subsection{Proof of CDF}
    According to the definition of CDF, we have
    \begin{equation}\label{CDFde}
    {F_{{\gamma _k}}}\left( \gamma  \right) = \int_0^\gamma  {{f_{{\gamma _k}}}\left( x \right){\rm{d}}x}.
    \end{equation}
    Combining \eqref{PDFFINAL} with \eqref{CDFde}, we have
    \begin{align}\label{CDFGO}
    &{F_{{\gamma _k}}}\left( \gamma  \right) = \frac{{{m_{fk}}^{{m_{fk}}}{{\left( {{m_{sk}} - 1} \right)}^{{m_{sk}}}}{{\bar z}_k}^{{m_{sk}}}}}{{{{\left( {{P_{Ik}}{\eta _k}} \right)}^{{N_{Ik}}}}{\Gamma\!\left( {{N_{Ik}}} \right)}{{\left( {{P_k}D_{{\rm{u}}k}^{ - {\alpha _k}}} \right)}^{{m_{fk}}}}B\!\left( {{m_{fk}},{m_{sk}}} \right)}}
    \notag\\&\times
    {\left( {\frac{1}{{2\pi i}}} \right)^{\rm{2}}}{\left( {\frac{{{P_k}D_{{\rm{u}}k}^{ - {\alpha _k}}}}{{{m_{fk}}}}} \right)^{{m_{fk}} + {m_{sk}}}}\frac{{{{\left( {{\sigma ^2}} \right)}^{{N_{Ik}} + {m_{fk}}}}}}{{\Gamma\!\left( {{m_{fk}} + {m_{sk}}} \right)}}
    \notag\\&\times
    \int_{{{\cal L}_1}} {\int_{{{\cal L}_2}} {\frac{{\Gamma\!\left( {{m_{sk}} - {s_1} - {N_{Ik}}} \right)\Gamma\!\left( { - {s_1}} \right)\Gamma\!\left( {{m_{fk}} + {m_{sk}} + {s_2}} \right)}}{{\Gamma\!\left( {{m_{sk}} + {s_2}} \right){\Gamma ^{ - 1}}\left( { - {s_2}} \right){\Gamma ^{ - 1}}\left( {{s_1} + {N_{Ik}} + {s_2}} \right)}}} } 
    \notag\\&\times
    {I_{\rm{2}}}{\left( {\frac{{{\sigma ^2}}}{{{P_{Ik}}{\eta _k}}}} \right)^{{s_1}}}{\left( { - {\Lambda _k}} \right)^{{s_2}}}{\rm{d}}{s_2}{\rm{d}}{s_1},
    \end{align}
    where the $I_2$ can be expressed as  
    \begin{equation}
    {I_{\rm{2}}}{\rm{ = }}\int_0^\gamma {\frac{{{x^{{m_{fk}} - 1}}}}{{{{\left( {{\Lambda _k} + x{\sigma ^2}} \right)}^{{m_{fk}} + {m_{sk}} + {s_2}}}}}{\rm{d}}x}.
    \end{equation}
    With the help of, we can solve $I_2$ as
    \begin{align}
    {I_{\rm{2}}}& = \frac{{{\gamma ^{{m_{fk}}}}}}{{{\Lambda _k}^{{m_{fk}} + {m_{sk}} + {s_2}}{m_{fk}}}}
    \notag\\ &\times
    F\left( {{m_{fk}} + {m_{sk}} + {s_2},{m_{fk}};1 + {m_{fk}}; - \frac{{{\sigma ^2}}}{{{\Lambda _k}}}\gamma } \right).
    \end{align}
    According to the integral expression of the hyper-geometric function \cite[eq. (9.113)]{gradshteyn2007}, we can substitute $I_2$ into \eqref{CDFGO} and obtain \eqref{CDFFINAL}, which completes the proof.

    \section{Proof of Lemma \ref{2lemma}}\label{2lemmaA}
    \renewcommand{\theequation}{B-\arabic{equation}}
    \setcounter{equation}{0}
    Let $t \triangleq x \rho$. We can re-write $I_A$ as
    \begin{align}
    &{I_A} = {\rho ^{ - b - c - 1}}\exp\!\left( {\frac{a}{d}} \right)
    \notag\\&\times
    \int_{\rho a}^\infty  {{t^b}{{\left( {t - \rho a} \right)}^c}\exp\!\left( { - \frac{t}{{\rho d}}} \right)F\left( {\alpha ,\beta ,\varepsilon ; - t} \right){\rm{d}}t} .
    \end{align}
    
    Because $\rho$ is small, we can further express $I_A$ as
    \begin{align}\label{jinsieq}
    {I_A} & \approx  {\rho ^{ - b - c - 1}}\exp\!\left( {\frac{a}{d}} \right)
    \notag\\&\times
    \int_0^\infty  {{t^{b + c}}\exp\!\left( { - \frac{t}{{\rho d}}} \right)F\left( {\alpha ,\beta ,\varepsilon ; - t}\right){\rm{d}}t}.
    \end{align}
    With the help of \cite[eq. (9.113)]{gradshteyn2007}, we obtain
    \begin{align}\label{aegaegr}
    {I_A} &= {\rho ^{ - b - c - 1}}\exp\!\left( {\frac{a}{d}} \right)\frac{{\Gamma\!\left( \varepsilon  \right)}}{{\Gamma\!\left( \alpha  \right)\Gamma\!\left( \beta  \right)}} \frac{1}{{2\pi i}}
    \notag\\&\times
    \int_{{{\cal L}_1}} {\frac{{\Gamma\!\left( {{s_1} + \alpha } \right)\Gamma\!\left( {{s_1} + \beta } \right)\Gamma\!\left( { - {s_1}} \right)}}{{\Gamma\!\left( {{s_1} + \varepsilon } \right)}}} {I_3}{\rm{d}}{s_1},
    \end{align}
    where
    \begin{equation}
    {I_3} = \int_0^\infty  {{t^{{s_1} + b + c}}\exp\!\left( { - \frac{t}{{\rho d}}} \right){\rm{d}}t}.
    \end{equation}
    By using \cite[eq. (3.351.3)]{gradshteyn2007} and \cite[eq. (8.339.1)]{gradshteyn2007}, $I_3$ can be solved as
    \begin{equation}
    {I_3} = {\left( {d \rho} \right)^{b + c + {s_1} + 1}}\Gamma\!\left( {1 + b + c + {s_1}} \right).
    \end{equation}
    Substituting $I_3$ into \eqref{aegaegr}, we have
    \begin{align}
    &{I_A} = \exp\!\left( {\frac{a}{d}} \right)\frac{{\Gamma\!\left( \varepsilon  \right){d^{b + c + 1}}}}{{\Gamma\!\left( \alpha  \right)\Gamma\!\left( \beta  \right)}} \frac{1}{{2\pi i}}
    \notag\\&\times
    \int_{{{\cal L}_1}} {\frac{{\Gamma\!\left( {{s_1} + \alpha } \right)\Gamma\!\left( {{s_1} + \beta } \right)\Gamma\!\left( { - {s_1}} \right)}}{{{\Gamma ^{ - 1}}\left( {1 + b + c + {s_1}} \right)\Gamma\!\left( {{s_1} + \varepsilon } \right)}}} {\left( {d\rho } \right)^{{s_1}}}{\rm{d}}{s_1}.
    \end{align}
    According to the definition of Meijer's $G$-function \cite[eq. (9.301)]{gradshteyn2007}, we can re-write $I_A$ as \eqref{APPFINAL} to complete the proof.
    
    \section{Proof of Theorem \ref{2them}}\label{2themA}
    \renewcommand{\theequation}{C-\arabic{equation}}
    \setcounter{equation}{0}
    Using the definition of CDF, we have
    \begin{equation}
    {F_{{\gamma _k}}}\left( \gamma  \right) = \int_0^\infty  {{F_U}\left( {\gamma x} \right){f_X}\left( x \right){\rm{d}}x},
    \end{equation}
    where 
    \begin{align}
    {F_U}\left( u \right) &= \Pr \left( {U < u} \right) = \Pr \left( {Z < \frac{u}{{{P_k}D_{{\rm{u}}k}^{ - {\alpha _k}}}}} \right) 
    \notag\\&
    = {F_Z}\left( {\frac{u}{{{P_k}D_{{\rm{u}}k}^{ - {\alpha _k}}}}} \right).
    \end{align}
    Thus, the CDF of $\gamma_k$ can be expressed as 
    \begin{align}\label{aegarebh}
    &{F_{{\gamma _k}}}\left( \gamma  \right) = \frac{1}{{{m_{fk}}B\!\left( {{m_{fk}},{m_{sk}}} \right)}}{\left( {\frac{\gamma }{{{\Lambda _k}}}} \right)^{{m_{fk}}}}\frac{1}{{{{\left( {{P_{Ik}}{\eta _k}} \right)}^{{N_{Ik}}}}{\Gamma\!\left( {{N_{Ik}}} \right)}}}
    \notag\\&\times
    \int_{{\sigma ^2}}^\infty  {{x^{{m_{fk}}}}{{\left( {x - {\sigma ^2}} \right)}^{{N_{I,k}} - 1}}\exp\!\left( { - \frac{{x - {\sigma ^2}}}{{{P_{Ik}}{\eta _k}}}} \right)} 
    \notag\\&\times
    F\left( {{m_{fk}},{m_{fk}} + {m_{sk}},{m_{fk}} + 1; - \frac{{\gamma x}}{{{\Lambda _k}}}} \right){\rm{d}}x.
    \end{align}
    The integration part in \eqref{aegarebh} can be solved with the help of Lemma \ref{2lemma}. Then, after some algebraic manipulations, we have
    \begin{align}\label{fajelkjf}
    &{F_{{\gamma _k}}}\left( \gamma  \right) = \frac{1}{{\Gamma\!\left( {{N_{Ik}}} \right)\Gamma\!\left( {{m_{sk}}} \right)\Gamma\!\left( {{m_{fk}}} \right)}}\exp\!\left( {\frac{{{\sigma ^2}}}{{{P_{Ik}}{\eta _k}}}} \right)
    \notag\\&\times
    \frac{1}{{2\pi i}}\int_{\cal L} {\frac{{\Gamma\!\left( {{s_2}} \right)\Gamma\!\left( {{s_2} + {m_{sk}}} \right)\Gamma\!\left( {{N_{Ik}} + {s_2}} \right)}}{{\Gamma\!\left( {{s_2} + 1} \right){\Gamma ^{ - 1}}\left( { - {s_2} + {m_{fk}}} \right)}}{{\left( {\frac{{\gamma {P_{Ik}}{\eta _k}}}{{{\Lambda _k}}}} \right)}^{{s_2}}}{\rm{d}}{s_2}}.
    \end{align}
    Using \cite[eq. (9.301)]{gradshteyn2007}, we can derive the CDF of $\gamma_k$ as \eqref{CDFFINAL}, which completes the proof.
    
      \section{Proof of Theorem \ref{berthm}}\label{berthmA}
    \renewcommand{\theequation}{D-\arabic{equation}}
    \setcounter{equation}{0}
    {\color{black}Using the definition of Gamma function \cite[eq. (8.350)]{gradshteyn2007}, we can express $E_k$ as
   \begin{equation}\label{geagaeg}
{E_k} = \frac{{{\color{black}{\lambda} _1}^{{\color{black}{\lambda} _2}}}}{{2\Gamma\!\left( {{\color{black}{\lambda} _2}} \right)}}\int_0^\infty  {{x^{{\color{black}{\lambda} _2} - 1}}{e^{ - {\color{black}{\lambda} _1}x}}{F_{{\gamma _k}}}\left( x \right){\rm{d}}x}.
\end{equation}
By substituting \eqref{fajelkjf} into \eqref{geagaeg}, we obtain
\begin{align}\label{egaeg3214}
&{E_k} = \frac{{{\color{black}{\lambda} _1}^{{\color{black}{\lambda} _2}}{\Gamma ^{ - 1}}\left( {{m_{sk}}} \right){\Gamma ^{ - 1}}\left( {{m_{fk}}} \right)}}{{2\Gamma\!\left( {{\color{black}{\lambda} _2}} \right)\Gamma\!\left( {{N_{Ik}}} \right)}}\exp\!\left( {\frac{{{\sigma ^2}}}{{{P_{I,k}}{\eta _k}}}} \right)\frac{1}{{2\pi i}}
\notag\\&\times
\int_{\cal L} {\frac{{\Gamma\!\left( {{s_2}} \right)\Gamma\!\left( {{s_2} + {m_{sk}}} \right)\Gamma\!\left( {{N_{Ik}} + {s_2}} \right)}}{{\Gamma\!\left( {{s_2} + 1} \right){\Gamma ^{ - 1}}\left( { - {s_2} + {m_{fk}}} \right)}}{{\left( {\frac{{{P_{I,k}}{\eta _k}}}{{{\Lambda _k}}}} \right)}^{{s_2}}}{I_4}{\rm{d}}{s_2}} ,
\end{align}
where 
\begin{equation}
{I_4}{\rm{ = }}\int_0^\infty  {{x^{{s_2} + {\color{black}{\lambda} _2} - 1}}{e^{ - {\color{black}{\lambda} _1}x}}{\rm{d}}x} .
\end{equation}
Using \cite[eq. (3.351.3)]{gradshteyn2007}, we can solve $I_4$ as
\begin{equation}
{I_4} = {\color{black}{\lambda} _1}^{ - {s_2} - {\color{black}{\lambda} _2}}\Gamma\!\left( {{s_2} + {\color{black}{\lambda} _2}} \right).
\end{equation}
Combining $I_4$ and \eqref{egaeg3214}, we derive $E_k$ as \eqref{BERFINAL} to complete the proof.}

    \end{appendices}
    \bibliographystyle{IEEEtran}
    \bibliography{IEEEabrv,Ref}
    \end{document}